%% file: main.tex
\documentclass[journal]{IEEEtran}
\ifCLASSINFOpdf
\else
\fi

\usepackage{stfloats}

\usepackage{amsmath,amssymb,amsfonts}
\usepackage{algorithm}
\usepackage{algorithmic}
\usepackage{graphicx}
\graphicspath{ {Figures/} }
\usepackage{stfloats}
\usepackage{multicol}
\usepackage{multirow}
\usepackage{tablefootnote}
\usepackage[flushleft]{threeparttable}
\usepackage{color, colortbl}

\newcommand{\Fp}{\mathbb{F}_p}

\definecolor{Gray}{gray}{0.9}

\usepackage{hyperref}
\hypersetup{
colorlinks=true,
linkcolor=blue,
filecolor=magenta,      
urlcolor=blue,
}

\begin{document}
%
\title{An Energy-Efficient Reconfigurable \\ DTLS Cryptographic Engine for \\ Securing Internet-of-Things Applications}
%
%
%

\author{
Utsav~Banerjee,
Andrew~Wright,
Chiraag~Juvekar,
Madeleine~Waller,
Arvind,
and~Anantha~P.~Chandrakasan%
\\ Dept. of EECS, Massachusetts Institute of Technology, Cambridge, MA, USA%
\thanks{\textcopyright $\,$ 2019 IEEE. Personal use of this material is permitted. Permission from IEEE must be obtained for all other uses, in any current or future media, including reprinting/republishing this material for advertising or promotional purposes, creating new collective works, for resale or redistribution to servers or lists, or reuse of any copyrighted component of this work in other works.}
\thanks{A revised version of this paper was published in the IEEE Journal of Solid-State Circuits (JSSC) - DOI: \href{https://dx.doi.org/10.1109/JSSC.2019.2915203}{10.1109/JSSC.2019.2915203}}
}

%
%

\markboth{}%
{Banerjee \MakeLowercase{\textit{et al.}}: An Energy-Efficient Reconfigurable DTLS Cryptographic Engine for Securing IoT Applications}
%



\maketitle

\begin{abstract}
This paper presents the first hardware implementation of the Datagram Transport Layer Security (DTLS) protocol to enable end-to-end security for the Internet of Things (IoT). A key component of this design is a reconfigurable prime field elliptic curve cryptography (ECC) accelerator, which is 238$\times$ and 9$\times$ more energy-efficient compared to software and state-of-the-art hardware respectively. Our full hardware implementation of the DTLS 1.3 protocol provides 438$\times$ improvement in energy-efficiency over software, along with code size and data memory usage as low as 8 KB and 3 KB respectively. The cryptographic accelerators are coupled with an on-chip low-power RISC-V processor to benchmark applications beyond DTLS with up to two orders of magnitude energy savings. The test chip, fabricated in 65 nm CMOS, demonstrates hardware-accelerated DTLS sessions while consuming 44.08 $\mu$J per handshake, and 0.89 nJ per byte of encrypted data at 16 MHz and 0.8 V.
\end{abstract}


%
\IEEEpeerreviewmaketitle

\input{body/01_intro.tex}

\input{body/02_overview.tex}
\input{body/03_riscv.tex}
\input{body/04_crypto.tex}
\input{body/05_dtls_engine.tex}
\input{body/06_measurements.tex}
\input{body/07_conclusion.tex}


%

\section*{Acknowledgment}

The authors would like to thank the Qualcomm Innovation Fellowship and Texas Instruments for funding this work, the TSMC University Shuttle Program for chip fabrication support, and Bluespec, Cadence and Mentor Graphics for providing CAD tools. They would also like to thank Mehul Tikekar, Priyanka Raina and Phillip Nadeau for their support during chip tape-out and testing.

\ifCLASSOPTIONcaptionsoff
  \newpage
\fi

\input{references.tex}


\end{document}

%% file: body/01_intro.tex
\section{Introduction}
\label{sec:intro}

\IEEEPARstart{T}{he} Internet of Things (IoT) is an ever-growing network of wireless electronic devices always connected to the Internet - collecting, processing and communicating data. While the IoT promises to enable fundamentally new applications, it is important to guarantee that the communication channel between each sensor node and the cloud server is secure, even in the presence of untrusted and potentially malicious network infrastructure \cite{keoh_iotsec_2014}. This is called end-to-end security, and protocols such as Datagram Transport Layer Security (DTLS) \cite{rescorla_tls_2018, rescorla_dtls_2018} enable the establishment of mutually authenticated confidential channels between IoT sensor nodes and the cloud. DTLS employs elliptic curve-based public key cryptographic techniques to authenticate the two end points and establish shared secret keys, which are then used to encrypt application data. TLS version 1.3 has recently been standardized by the Internet Engineering Task Force (IETF), and is considered to be one of the most suited protocols for securing the IoT \cite{keoh_iotsec_2014}. While this makes DTLS an ideal solution for IoT, the associated computational cost makes software-only implementations prohibitively expensive for resource-constrained embedded devices \cite{banerjee_eedtls_2017}. IoT devices are usually powered by batteries, which are expected to last several years, or through energy harvesting. Moreover, commercially available IoT platforms use micro-controllers with limited instruction and data memory. Therefore, it is essential to have a DTLS implementation which not only has minimal energy consumption but also comes with a small memory footprint. To address these challenges, we present the first hardware implementation \cite{banerjee_isscc_2018} of DTLS 1.3, based on version 18 of the protocol draft \cite{rescorla_dtls_2018}. Our reconfigurable elliptic curve cryptography (ECC) accelerator enables two orders of magnitude energy savings, while a dedicated DTLS engine offloads protocol control flow to hardware reducing program code and memory usage by an order of magnitude. An on-chip RISC-V processor exercises the flexibility of the cryptographic accelerators to demonstrate security applications beyond DTLS.

An overview of the DTLS protocol is presented in Section \ref{sec:overview}, along with our high-level system architecture. Section \ref{sec:riscv} describes the RISC-V processor, Section \ref{sec:crypto} provides architectural details of the energy-efficient cryptographic primitives and Section \ref{sec:dtls_engine} describes the design of the DTLS engine. Measurement results from the test chip are presented in Section \ref{sec:results}, and Section \ref{sec:conclusion} provides concluding remarks.

%% file: body/02_overview.tex
\section{System Architecture}
\label{sec:overview}

\subsection{Transport Layer Security}

\begin{figure}[!t]
\centering
\includegraphics[width=3.4in]{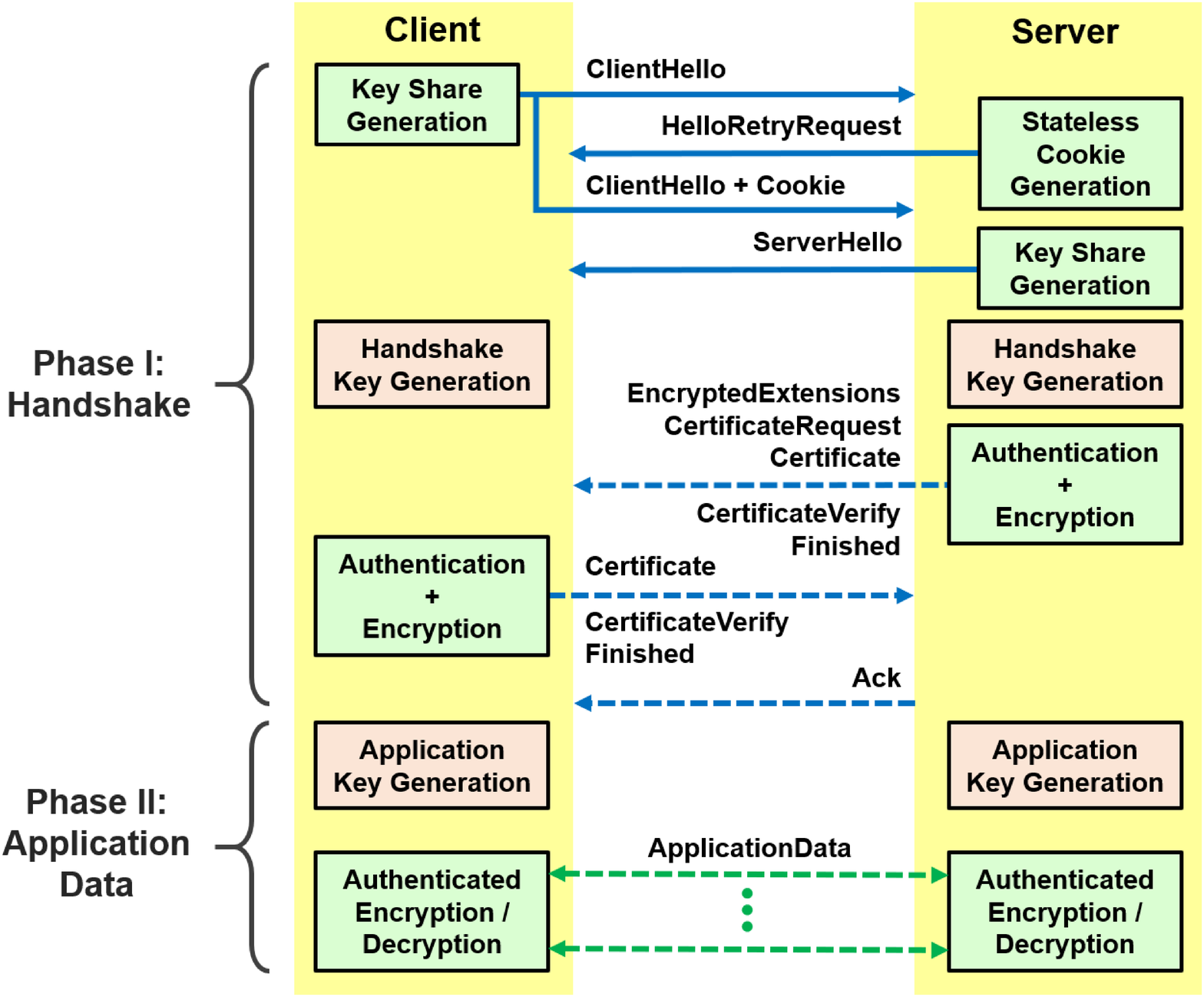}
\caption{Overview of the DTLS handshake protocol with digital certificate-based mutual authentication and key exchange (dashed arrows indicate that the messages are encrypted).}
\label{dtls_handshake}
\end{figure}

The DTLS protocol can be divided into two major phases - \textit{handshake} and \textit{application data} (Fig. \ref{dtls_handshake}). The handshake starts with the client (sensor node) and the server agreeing upon protocol parameters such as the cryptographic algorithms to be used. Next, a Diffie-Hellman key exchange \cite{menezes_handbook_1996} is performed to establish a shared secret over the untrusted channel. The subsequent handshake messages are completely encrypted using keys derived from this shared secret. Following this, the client and the server authenticate each other through digital certificate verification. Finally, the two parties verify the integrity of the information exchanged in the above steps, to prevent man-in-the-middle attacks. At this point, a mutually authenticated confidential channel has been established between the client and the server. This channel can then be used, in the application data phase, to exchange data encrypted under a new set of keys derived from the handshake parameters.

The DTLS specification lists a set of recommended cryptographic algorithms, also known as \textit{cipher suites}, to be used for performing the handshake and encrypting data. In this work, we consider DTLS connections implementing the TLS\_ECDHE\_ECDSA\_WITH\_AES\_128\_GCM\_SHA256 cipher suite, where elliptic curve cryptography \cite{hankerson_ecc_2004} is used for endpoint authentication and key exchange, AES-128-GCM (Advanced Encryption Standard in Galois/Counter Mode) \cite{nist_aes_2001, nist_gcm_2007} is used for authenticated encryption, and SHA2-256 (Secure Hash Algorithm 2) \cite{nist_sha2_2012} is used for message hashing, key derivation and pseudo-random number generation. The handshake phase involves $\approx 100$ invocations each of the AES-GCM and SHA primitives, which operate in blocks of 128 bits and 512 bits respectively; one ECDHE (Elliptic Curve Diffie-Hellman Key Exchange), and at least two ECDSA (Elliptic Curve Digital Signature Algorithm) operations (one ECDSA-Sign and at least one ECDSA-Verify). Once the handshake is complete, encryption or decryption of application data requires one invocation of AES-GCM per 128-bit block of data.

While the computation energy spent during each DTLS handshake is constant for a given cipher suite, the energy required during the application data phase is a direct function of the application payload size. Let us denote the handshake energy and the encrypted application data energy per byte of payload as $E_{handshake}$ and $E_{appdata}$ respectively, the session duration (time interval between two consecutive handshakes) as $t_{session}$ and the application data period (time interval between two consecutive application data transmissions) as $t_{appdata}$. Then, for $N$ bytes of application payload, the total computation energy during a session is given by
\begin{align*}
E_{total} &= E_{handshake} + (N \times \frac{t_{session}}{t_{appdata}} \times E_{appdata})
\end{align*}
since the total number of data transmissions during a session is $t_{session}/t_{appdata}$. The fraction of energy spent in handshake computations is $E_{handshake}/E_{total}$. The session duration $t_{session}$ is dictated by security requirements of the application -- more frequent handshakes (to establish new session keys), that is, smaller $t_{session}$, imply stronger security guarantees, e.g., medical devices authenticate more often than industrial sensors. The application data rate is calculated as $N/t_{appdata}$, which also depends on the application, e.g., industrial sensors typically send small packets of data every hour while medical devices send large amounts of data every minute or every second.

\begin{figure}[!t]
\centering
\includegraphics[width=3.4in]{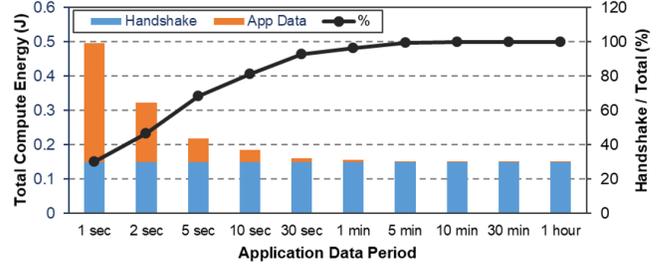}
\caption{DTLS computation energy breakdown and percentage of total compute energy spent in handshake, for $N = 32$ bytes of application payload, session duration $t_{session} = 1$ day and varying application data period $t_{appdata}$.}
\label{dtls_compute_breakdown}
\end{figure}

\begin{figure}[!t]
\centering
\includegraphics[width=3.4in]{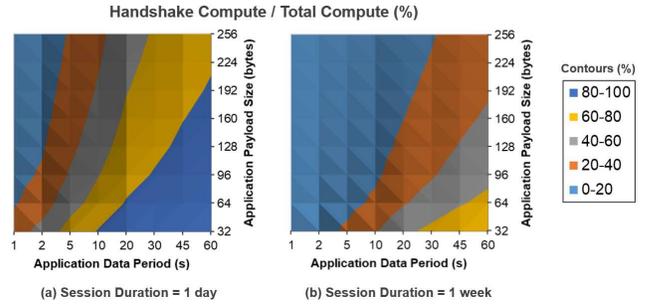}
\caption{Contour plots showing the percentage of total compute energy spent in handshake, for varying application payload size $N$ and varying application data period $t_{appdata}$, for session duration of (a) 1 day and (b) 1 week.}
\label{dtls_compute_contours}
\end{figure}

To understand the effect of application data rate on compute energy, we consider $E_{handshake} = 150$ mJ and $E_{appdata} = 125$ nJ as measured from an embedded software implementation of DTLS \cite{banerjee_eedtls_2017}. For devices handshaking once every day and payload size of $N = 32$, the breakdown of computation energy is shown in Fig. \ref{dtls_compute_breakdown}. We observe that the percentage of energy spent in DTLS handshake is around 30\% when data is transmitted every second, and more than 99\% when data is transmitted every hour. To further analyze the effects of these parameters, contour plots are shown in Fig. \ref{dtls_compute_contours} for $t_{session} = 1$ day and $t_{session} = 1$ week. As expected, the handshake energy becomes a larger fraction of total energy for smaller $N$, larger $t_{appdata}$ and smaller $t_{session}$. We observe that the total computation energy for a software implementation of DTLS is of the order of 0.1-0.5 J, which is dominated by either handshake computations or application data encryption depending on the application parameters. Therefore, it is essential to design energy-efficient hardware to accelerate both handshake and application data computations for low-power IoT devices secured by DTLS.

\subsection{Chip Overview}

\begin{figure}[!t]
\centering
\includegraphics[width=2.5in]{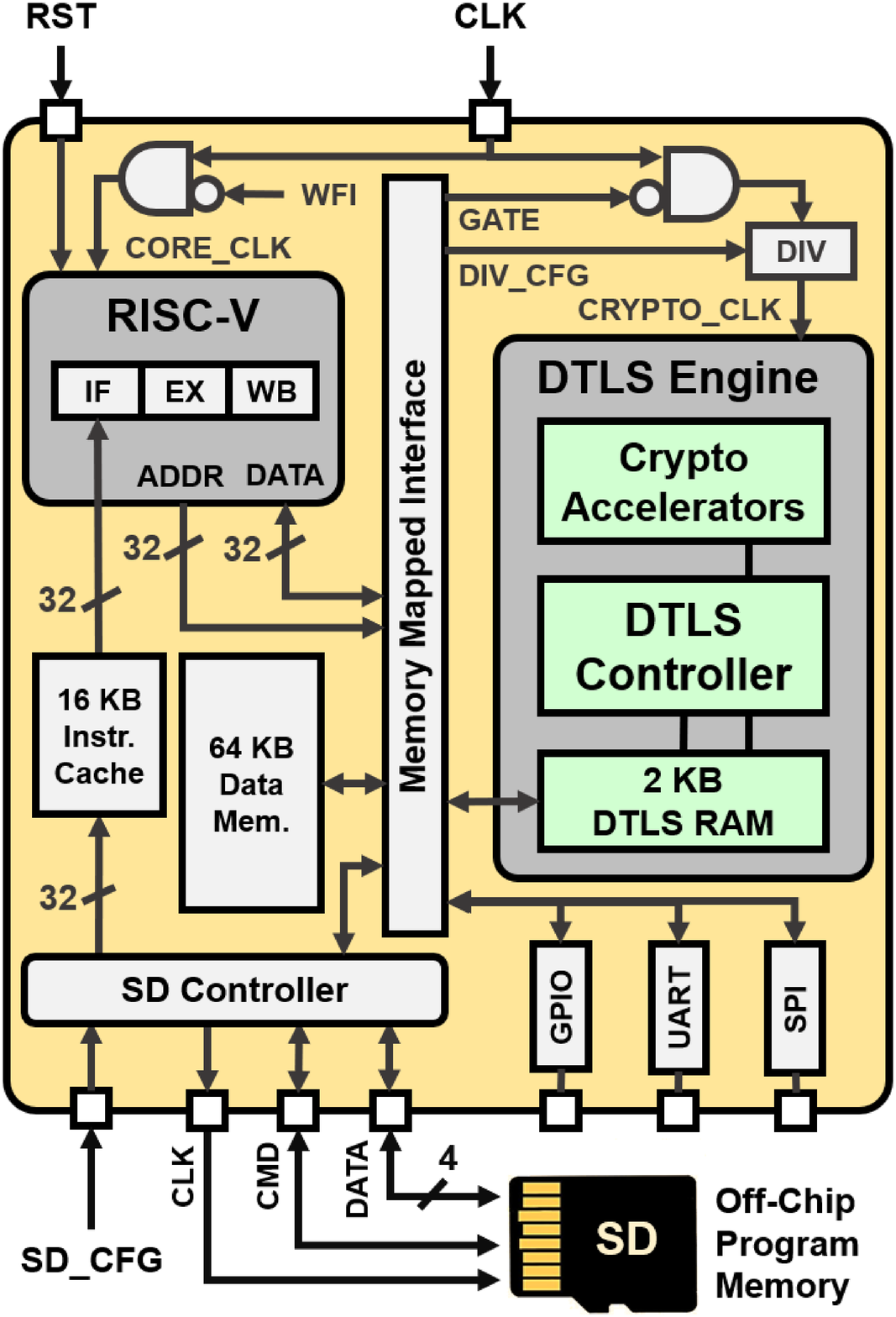}
\caption{System block diagram with an overview of the hardware modules implemented on the test chip.}
\label{chip_overview}
\end{figure}

Fig. \ref{chip_overview} shows the system block diagram. It consists of a 3-stage (IF: instruction fetch, EX: execute, WB: write back) RISC-V processor \cite{waterman_riscv_2014} supporting the RV32I instruction set, with 16 KB instruction cache and 64 KB data memory, and an SD (Secure Digital) card used as backing store for larger programs. A DTLS engine (DE), comprised of a protocol controller, a dedicated 2 KB RAM, and AES-128-GCM, SHA2-256 and prime field ECC primitives, accelerates both the handshake and application data phases of the DTLS protocol. Sleep mode is implemented on the RISC-V, to save power, by gating its clock when cryptographic tasks are delegated to the DE. The DE uses a dedicated hardware interrupt to wake the processor on completion of these tasks. The DE is clocked by a software-controlled divider to decouple the processor operating frequency from the long critical paths in the ECC accelerator. A memory-mapped interface provides access to the DTLS engine, through the DTLS RAM, not only for executing DTLS protocol workloads but also for standalone computations in the cryptographic accelerators. The same interface is also used to communicate with peripherals such as GPIO (General Purpose Input / Output), UART (Universal Asynchronous Receiver / Transmitter) and SPI (Serial Peripheral Interface) through RISC-V software. This memory-mapped interface, along with the accelerator interrupts, behaves very similarly to the Rocket Custom Coprocessor (RoCC) interface used by the Rocket RISC-V core \cite{asanovic_rocket_2016} to interface with accelerators. However, the memory-mapped approach does not require building custom instructions, thus simplifying the software tool-chain.

%% file: body/03_riscv.tex
\section{RISC-V Microprocessor}
\label{sec:riscv}

The RISC-V processor on the test chip is a 32-bit core, designed in Bluespec System Verilog, supporting the integer subset of instructions (RV32I) with user and machine privilege modes. The RISC-V core was designed to not only seamlessly interface with the DTLS engine but also efficiently implement the DTLS protocol in software. In order to support the large instruction storage required by the DTLS software (detailed in Section \ref{sec:results}), an SD card is used as off-chip program memory with the processor instruction cache reading program blocks from the card through an on-chip SD controller.

The instruction cache, being backed by an SD card, has to deal with a larger memory access granularity (512 bytes vs 64 bytes) and a longer memory access latency compared to typical microprocessor caches backed by DRAM modules. To match the access granularity of the SD card, the instruction cache was designed with a block size of 512 bytes, and the cache was made 4-way set associative with a tree-based pseudo-LRU (least recently used) replacement policy to reduce the number of cache misses. Apart from the 16 KB SRAM for storing instruction words, the cache also contains an 84-byte register array for storing tags. The tag array accesses only a single way's tag at a time, instead of all four tags, and the 16 KB SRAM only accesses one 32-bit instruction at a time, thus reducing access power. To reduce the overhead of accessing one way at a time, the cache has an MRU (most recently used) way predictor to estimate which way will be used for each instruction fetch. This way predictor reuses the meta-data from the replacement policy to determine the most recently used way. To further reduce the cache access penalty, a register is used to cache the last tag read from the tag array so that the tag array is accessed only when switching cache lines.

\begin{table}[!b]
\renewcommand{\arraystretch}{1.2}
\caption{Comparison of our RISC-V core with state of the art}
\label{table:riscv_comparison}
\centering
\begin{tabular}{|l|c|c|c|c|}
\hline
\rowcolor{Gray}
\textbf{Design} & \textbf{Arch} & \textbf{Tech} & \textbf{Voltage} & \textbf{Energy} \\
\rowcolor{Gray}
 & & \textbf{(nm)} & \textbf{(V)} & \textbf{(pJ / cyc)} \\
\hline
\hline
Duran \textit{et al.}, & RISC-V & \multirow{2}{*}{130} & \multirow{2}{*}{1.2} & \multirow{2}{*}{167} \\
LASCAS 2017 \cite{duran_riscv_2017} & RV32IM & & & \\
\hline
Uytterhoeven \textit{et al.}, & RISC-V & \multirow{2}{*}{28} & \multirow{2}{*}{0.38} & \multirow{2}{*}{8.81} \\
ESSCIRC 2018 \cite{dehaene_riscv_2018} & RV32IM & & & \\
\hline
\multirow{2}{*}{\textbf{This work}} & RISC-V & \multirow{2}{*}{65} & \multirow{2}{*}{0.8} & \multirow{2}{*}{40.36} \\
& RV32I & & & \\
\hline
\end{tabular}
\end{table}


The SD controller is designed to reduce miss latency by using the SD bus protocol \cite{sd_phy_2017} where card data is accessed 4 bits at a time, instead of the bit-serial SPI mode. Both SDHC and SDXC cards are supported, with clock frequencies up to 25 MHz. The SD clock is generated from the system clock through a clock divider configured externally (through SD\_CFG). Memory-mapped SPI and GPIO peripherals on the chip are used to interface with off-chip components. The UART peripheral, driven by a custom software library, is used to send debug messages to the host computer during testing.

\begin{figure}[!t]
\centering
\includegraphics[width=3.4in]{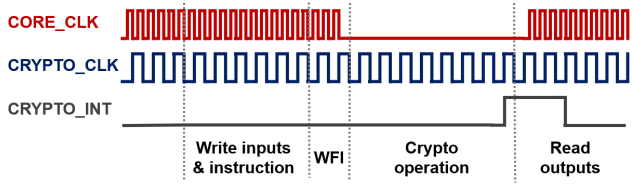}
\caption{Processor clock gating during WFI (wait for interrupt).}
\label{riscv_wfi}
\end{figure}

The RISC-V core is also equipped with an interrupt controller to handle interrupts from the cryptographic accelerators, the peripherals as well as off-chip. The interrupts can be individually enabled and programmed to be edge or level triggered through software. Executing the \textit{wait for interrupt} (WFI) instruction gates the clock feeding the RISC-V core and its instruction cache and data memory, as shown in Fig. \ref{riscv_wfi}, which enables power savings when the DTLS engine is accelerating cryptographic computations. The interrupt controller wakes them up when the appropriate interrupt is received.

When executing the Dhrystone benchmark, our RISC-V processor consumes 40.36 $\mu$W/MHz at 0.8 V, and achieves 0.96 DMIPS/MHz which is comparable to the ARM Cortex-M0 processor \cite{arm_cortexm}. Table \ref{table:riscv_comparison} compares the energy-efficiency of our design with some recent embedded-scale processor implementations.

%% file: body/04_crypto.tex
\section{Cryptographic Primitives}
\label{sec:crypto}

As discussed in Section \ref{sec:overview}, DTLS requires not only symmetric cryptography primitives such as AES and SHA, but also public key protocols using ECC. In this section, we provide details of the energy-efficient implementations of these primitives, including architectural optimizations, design space exploration and on-chip characterization results.

\subsection{AES in Galois/Counter Mode (AES-GCM)}

\begin{table*}[!b]
\renewcommand{\arraystretch}{1.2}
\caption{Comparison of our AES-128 with state of the art}
\label{table:aes_comparison}
\centering
\begin{tabular}{|l|c|c|c|c|c|c|}
\hline
\rowcolor{Gray}
\textbf{Design} & \textbf{Tech} & \multicolumn{2}{c|}{\textbf{Area}} & \textbf{Cycles} & \textbf{Voltage} & \textbf{Energy} \\ \cline{3-4}
\rowcolor{Gray}
 & \textbf{(nm)} & \textbf{(mm$^2$)} & \textbf{(kGE)} & \textbf{/ Block} & \textbf{(V)} & \textbf{(pJ / bit)} \\
\hline
\hline
Hamalainen \textit{et al.}, EUROMICRO 2006 \cite{hamalainen_aes_2006} \textsuperscript{a} & 130 & - & 3.2 & 160 & 1.2 & 37.5 \\
\hline
\multirow{2}{*}{Mathew \textit{et al.}, JSSC 2011 \cite{mathew_aes_2011}} & \multirow{2}{*}{45} & \multirow{2}{*}{0.15} & \multirow{2}{*}{-} & \multirow{2}{*}{5} & 1.1 & 2.3 \\ \cline{6-7}
 & & & & & 0.32 & 0.5 \\
\hline
\multirow{2}{*}{Mathew \textit{et al.}, JSSC 2015 \cite{mathew_aes_2015}} & \multirow{2}{*}{22} & \multirow{2}{*}{0.0022} & \multirow{2}{*}{1.9} & \multirow{2}{*}{336} & 0.9 & 30.1 \\ \cline{6-7}
 & & & & & 0.34 & 5.9 \\
\hline
\multirow{2}{*}{Zhang \textit{et al.}, VLSIC 2016 \cite{zhang_aes_2016}} & \multirow{2}{*}{40} & \multirow{2}{*}{0.0043} & \multirow{2}{*}{2.3} & \multirow{2}{*}{336} & 0.9 & 8.9 \\ \cline{6-7}
 & & & & & 0.47 & 2.2 \\
\hline
\textbf{This work} & 65 & 0.015 \textsuperscript{b} & 10.6 \textsuperscript{b} & 11 & 0.8 & 4.08 \textsuperscript{c} \\
\hline
\multicolumn{7}{l}{\textsuperscript{a} Post-synthesis area and power reported in \cite{hamalainen_aes_2006} \textsuperscript{b} Area of final placed-and-routed design \textsuperscript{c} Measured energy}
\end{tabular}
\end{table*}

The DTLS protocol uses AES-128 in the GCM mode for authenticated encryption with associated data (AEAD), that is, it simultaneously guarantees confidentiality, integrity, and authenticity of the data. The AES-128 cipher uses 128-bit keys to encrypt 128-bit plain-text blocks over 10 iteration rounds, with each round performing a set of linear and non-linear transformations on the cipher's internal state. The S-Box is the most important non-linear component of AES, used both in encryption and key expansion. In this work, we have used the low-power low-area S-Box design proposed in \cite{canright_sbox_2004}.


\begin{figure}[!t]
\centering
\includegraphics[width=2.5in]{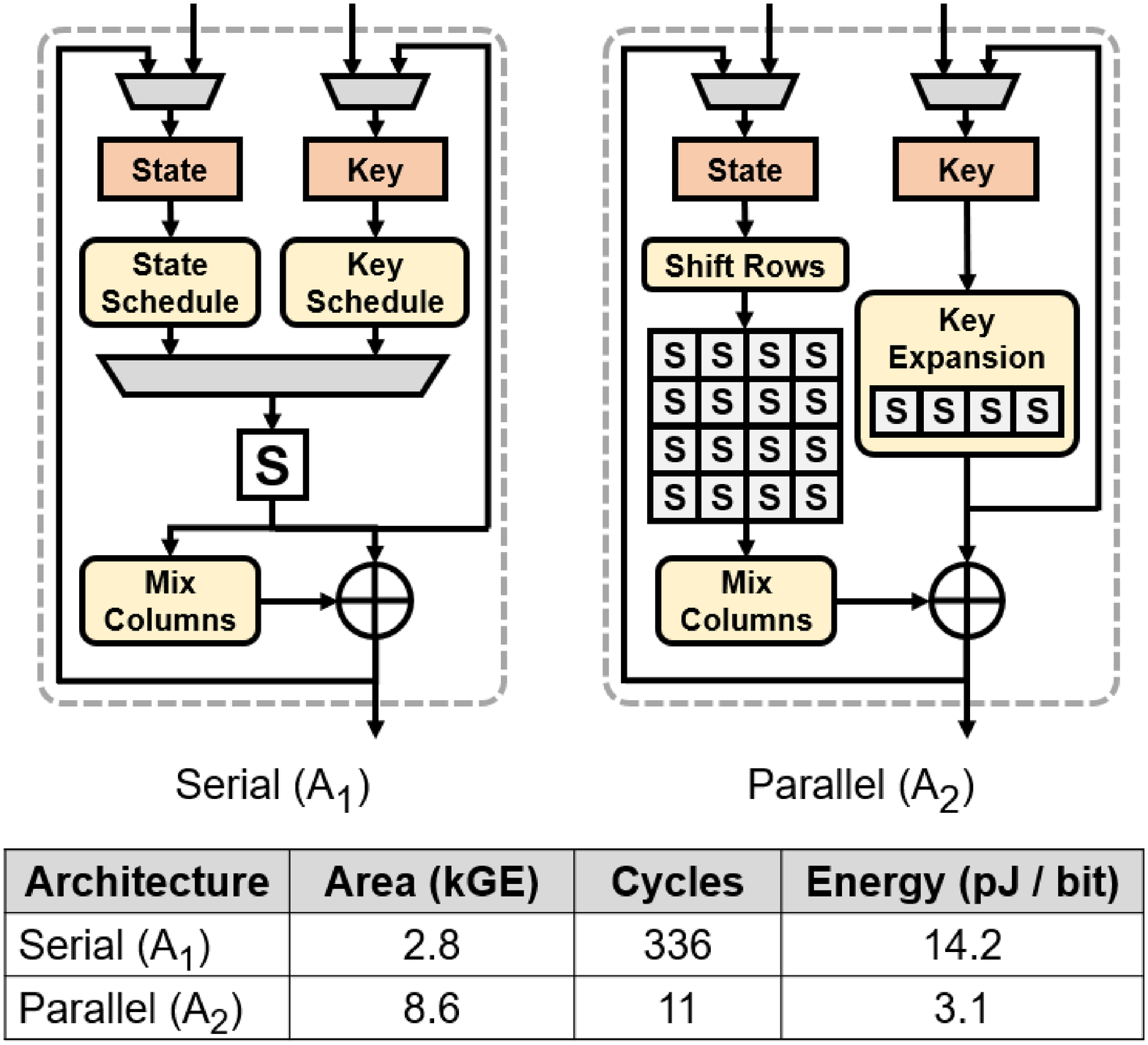}
\caption{Comparison of AES architectures - $A_1$: serial and $A_2$: parallel.}
\label{aes_arch}
\end{figure}

\begin{figure}[!t]
\centering
\includegraphics[width=3.4in]{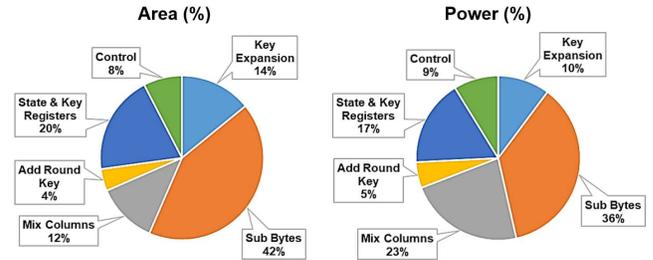}
\caption{Simulated area and power breakdown of the 128-bit data-path 11-cycle AES design.}
\label{aes_area_power_breakdown}
\end{figure}

To explore the effects of AES data-path size on area and energy-efficiency, we implemented two different AES architectures, as shown in Fig. \ref{aes_arch}:
\begin{itemize}
\item $A_1$, with 8-bit data-path and one S-Box, processes the state and the round key on separate cycles 8 bits at a time, and takes 336 cycles to encrypt a block.
\item $A_2$, with 128-bit data-path and 20 S-Boxes, processes the state and the round key together in a single cycle, and takes 11 cycles to encrypt a block.
\end{itemize}
The 8-bit architecture $A_1$ replicates the optimizations proposed in \cite{mathew_aes_2015} and \cite{zhang_aes_2016} to reduce the number of temporary registers. Fig. \ref{aes_arch} compares the area, performance and energy-efficiency of the two designs, as determined from post-synthesis simulations in 65 nm LP process at 1.2 V. The 128-bit parallel design $A_2$ is $4.6\times$ more energy-efficient and $30\times$ faster, at the cost of $3\times$ increase in logic area. Fig. \ref{aes_area_power_breakdown} shows the breakdown of area and power consumption of different components of this design. The S-Boxes account for a large fraction of area and power, which reaffirms our choice of composite-field S-Box.

\begin{figure}[!t]
\centering
\includegraphics[width=3.4in]{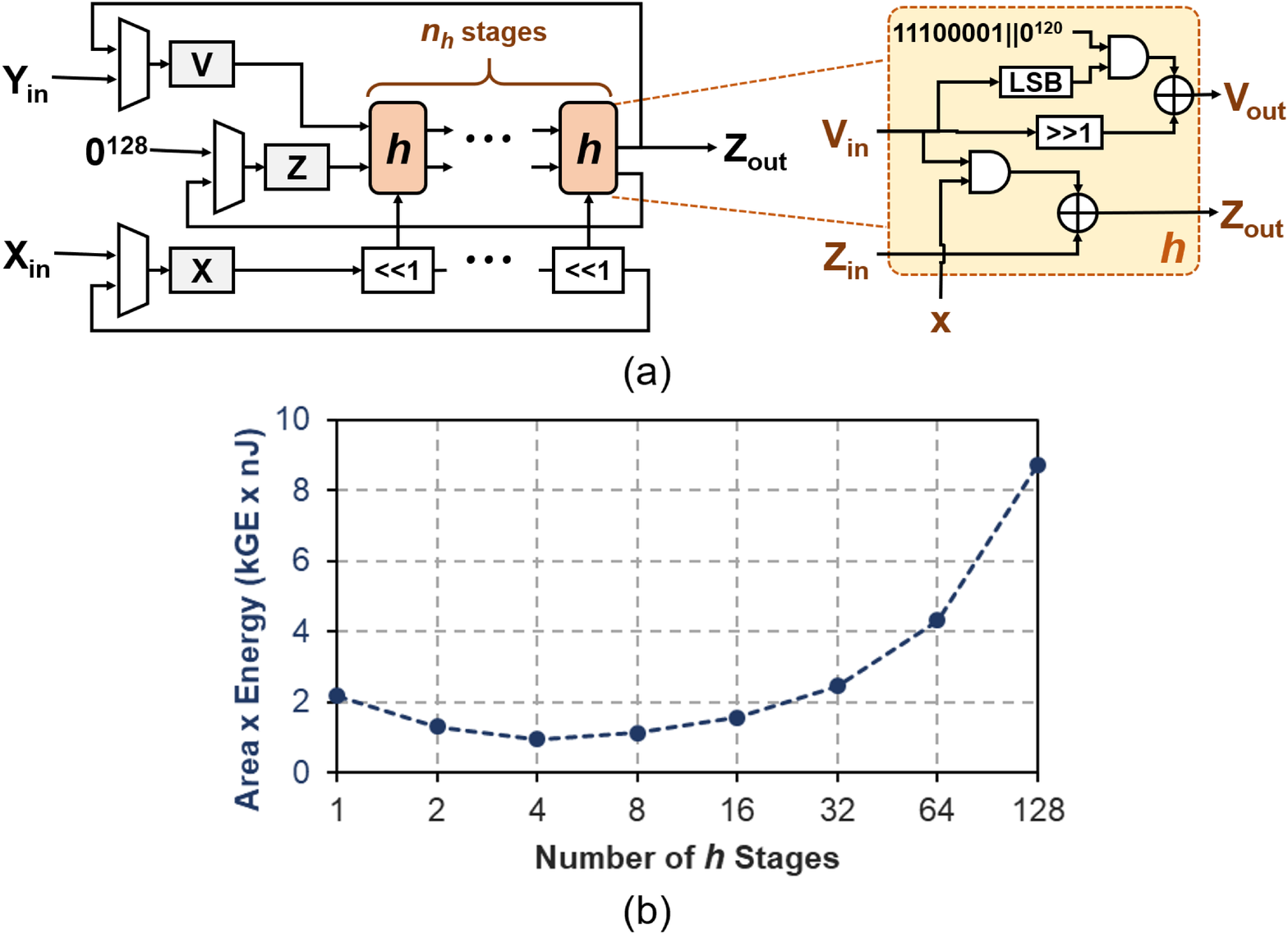}
\caption{(a) Implementation of GHASH Galois multiplier in hardware and (b) effect of number of multiplier stages ($n_h$) on area and energy.}
\label{aes_ghash}
\end{figure}

Table \ref{table:aes_comparison} compares our AES-128 design with state of the art, both in terms of area and energy. Our design is smaller than the 128-bit data-path 2-stage pipelined AES design in \cite{mathew_aes_2011}, while having comparable energy consumption, after accounting for voltage and technology scaling. In comparison to \cite{hamalainen_aes_2006, mathew_aes_2015, zhang_aes_2016}, which are all 8-bit data-path serial implementations, our design is more energy-efficient, when accounting for voltage and technology scaling, but at the cost of larger area. Note that our AES could not be characterized at voltages smaller than 0.8 V because all logic and SRAMs on our chip are powered by a single supply rail. Also, our measured AES energy includes leakage from the entire chip, since other components were clock-gated but not power-gated.


AES-GCM uses the AES forward cipher for both encryption and decryption, and a Galois multiplication-based special hash function called \textit{GHASH} for authentication \cite{nist_gcm_2007}. AES-GCM employs the counter mode of operation, which concatenates a counter value with the initialization vector $IV$, and encrypts it with the secret key using AES. The result of this encryption is then XOR-ed with the plain-text to generate the cipher-text. Like all counter modes, this essentially acts as a stream cipher, therefore it is important to ensure that a different $IV$ is used for each stream that is encrypted.

The Galois multiplier in GHASH can be implemented in hardware using one or more copies of the basic function which we denote as $h$: $Z_{i+1}$ = $Z_i \oplus x_i \cdot V_i$ and $V_{i+1}$ = $(V_i >> 1)$ $\oplus$ \{LSB($V_i$)\} $\cdot$ ($11100001||0^{120}$), as shown in Fig. \ref{aes_ghash}a. A Galois multiplier with $n_h$ stages requires $128/n_h$ cycles per multiplication, and the number of $h$-stages directly affects area, cycles per operation and energy consumption. Multiple Galois multipliers were synthesized to determine a suitable architecture, and their area-energy products were plotted as a function of the number of $h$-stages, as shown in Fig. \ref{aes_ghash}b. We observed that a 32-cycle design, with $n_h = 4$, has the lowest area-energy product, hence this version was used in our AES-GCM implementation. Since AES-GCM involves computing GHASH on the cipher-text, our design performs encryption and Galois multiplications in parallel, at 32 cycles per 128-bit data block. For $m$ blocks of associated data and $n$ blocks of plain-text (cipher-text), it takes $54 + 32 \cdot (m + n)$ cycles to encrypt (decrypt) and generate (verify) the GCM tag, where the fixed 54-cycle overhead accounts for computing the hash key, hashing the data length, computing the tag as well as configuring the key, IV and other encryption parameters. The final placed-and-routed design occupies 29.9 kGE area, including the 10.6 kGE AES, of which about 25\% is attributed to registers used to store input/output data, keys, intermediate states and configuration values. Energy consumption of our design is 11.88 pJ/bit at 0.8 V.

\subsection{Secure Hash Algorithm (SHA2)}

The SHA2-256 hash algorithm compresses messages of arbitrary lengths ($<2^{64}$ bits) and generates a unique 256-bit message digest. Since SHA2-256 operates on 512-bit blocks, the input message is padded to a multiple of 512 bits. The internal state of the hash function is initialized according to the SHA2 specification \cite{nist_sha2_2012}. The \textit{Message Schedule} takes 512-bit blocks of the padded message and sends 32-bit words $W_t$ to the main SHA2-256 \textit{Round} function, along with a round constant $K_t$. Each 512-bit block is digested over 64 iterations of the round function, and the state is updated. This continues till the entire message has been processed, and the final value of the state is the message digest.

\begin{figure}[!b]
\centering
\includegraphics[width=3.4in]{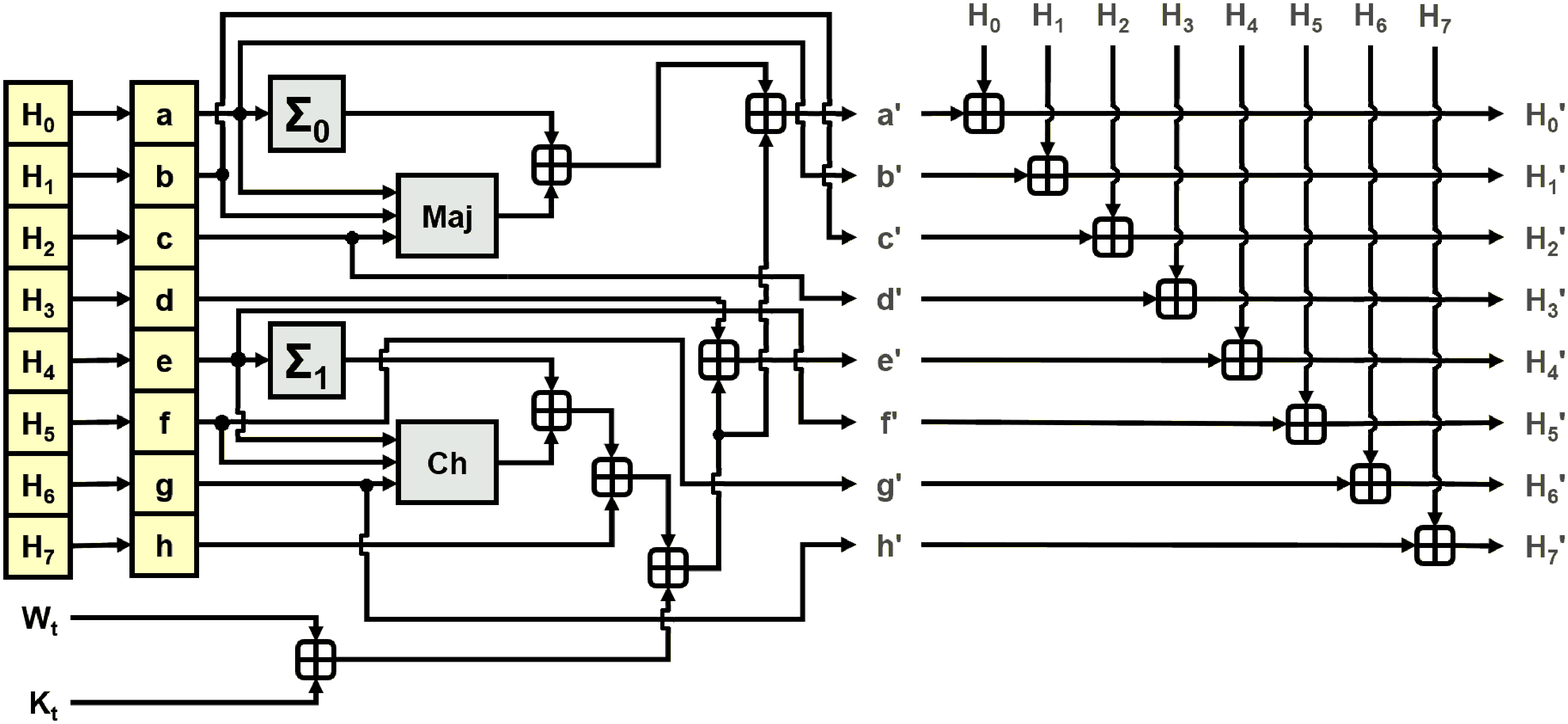}
\caption{Implementation of SHA2-256 round function in hardware.}
\label{sha2_round}
\end{figure}

Fig. \ref{sha2_round} shows details of the round function. The internal state consists of 16 32-bit registers $H_0 - H_7$ and $a - h$. The $\Sigma _0$, $\Sigma _1$, $Maj$ and $Ch$ functions are specified in \cite{nist_sha2_2012}, while $\boxplus$ denotes 32-bit addition modulo $2^{32}$, that is, the final carry is ignored. $H_0' - H_7'$ and $a' - h'$ denote the updated state values after one iteration. Although the state of the hash function is defined by $H_0 - H_7$, $a - h$ and the message schedule, we note that $H_0 - H_7$ completely define the SHA2-256 state after every 64 iterations of the round, that is, after every 512-bit block has been processed. This property can be exploited to implement efficient running hashes, as will be discussed in Section \ref{sec:dtls_engine}.

The critical paths in the round function were implemented using a combination of carry-save and ripple-carry adders to reduce latency. Messages are sent to the SHA2 core one byte at a time, and a counter is used to track the input data length, which is used by the SHA2 core to perform message padding. The SHA2-256 core computes $a'-h'$ in parallel to achieve increased energy-efficiency. Our final design occupies 18.2 kGE, and takes 65 cycles to process a 512-bit input block, while consuming 4.43 pJ/bit at 0.8 V.


\subsection{Reconfigurable Prime Field ECC}

Elliptic curve cryptography (ECC) is used in DTLS for both key exchange and digital signature protocols. We consider two types of elliptic curves over finite fields $\Fp$ of large prime characteristic $p$ -- short Weierstrass curves ($y^2 = x^3 + ax + b$) and Montgomery curves ($by^2 = x^3 + ax^2 + x$). All other prime curves (for $p \ne 2, 3$) can be transformed into the short Weierstrass form with a simple change of variables \cite{hankerson_ecc_2004}. ECC-based protocols can choose from a large set of standard curves, e.g., NIST, Curve25519, Brainpool, SEC and ANSSI. While existing literature in ECC hardware mostly focus on implementing a single family of curves \cite{hutter_sardines_2011, ingrid_tinyecc_2013, hutter_tags_2014, hutter_nacl_2015}, a similar approach is not suitable for DTLS because the standard allows a much wider choice of curves. This provides the motivation for our reconfigurable prime field ECC design, and we support curves over any prime up to 256 bits, which correspond to at most 128 bits of security.

\begin{figure}[!t]
\centering
\includegraphics[width=3.2in]{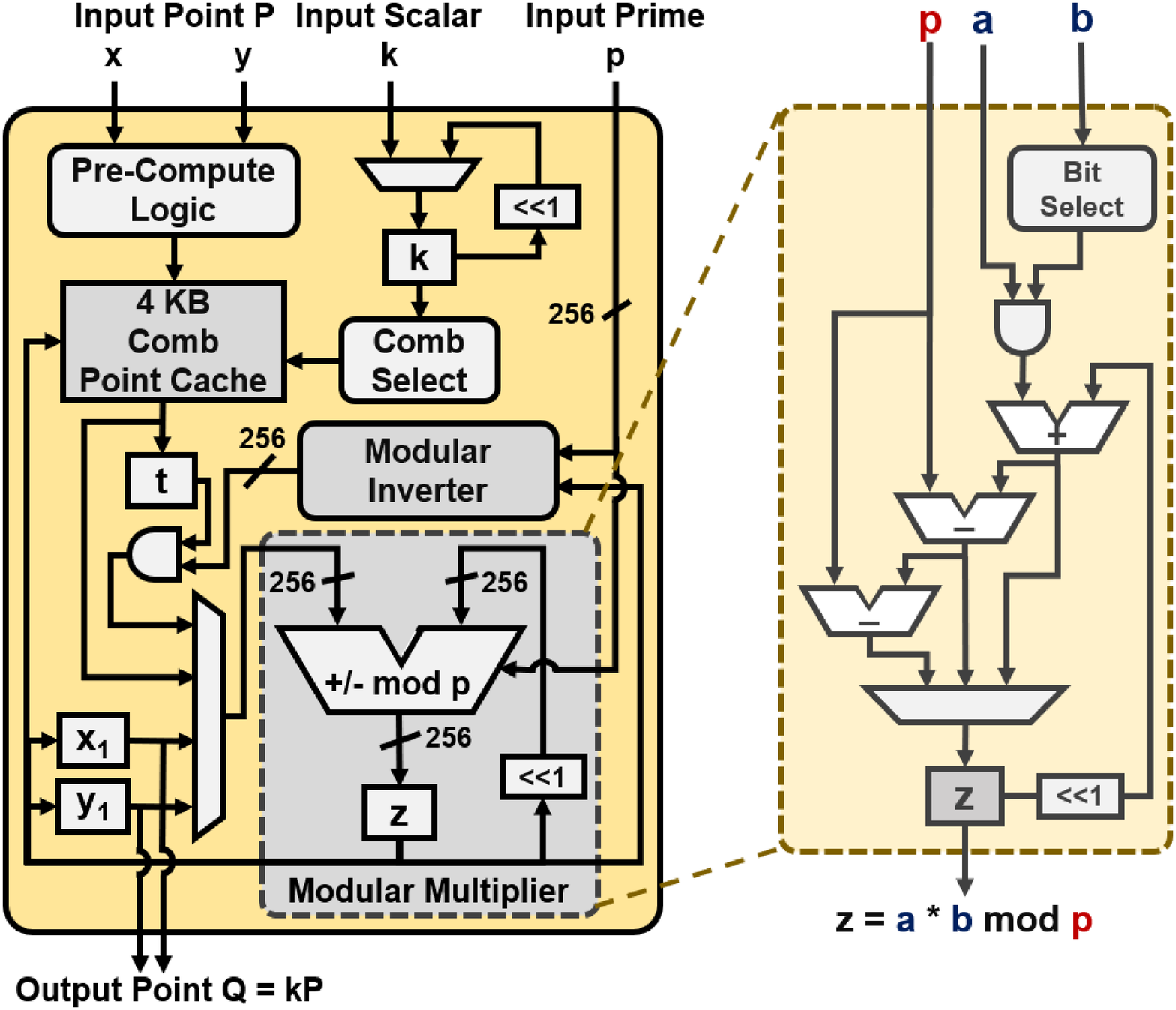}
\caption{Block diagram of the reconfigurable prime-field elliptic curve cryptography accelerator, along with detailed architecture of the modular multiplier implementing interleaved modular reduction.}
\label{ecc_block_diagram}
\end{figure}

The fundamental operations used in ECC are \textit{point addition} ($R = P + Q$), and \textit{point doubling} ($R = P + P$). Repeated additions of a point $P$ with itself is called ``elliptic curve scalar multiplication'' (ECSM). For any scalar $k$, the multiple $kP$ is computed as a series of point doubling (DBL) and point addition (ADD) operations, which can be decomposed into arithmetic in the finite field $\Fp$. This makes efficient modular arithmetic integral to both software and hardware implementations of ECC. Fig. \ref{ecc_block_diagram} describes our energy-efficient ECSM hardware, which can be configured with prime $p$ of variable length $t$ (up to 256 bits) and curve parameters $a$ and $b$. Given scalar $k$ and point $P (x, y)$, it generates $Q = kP$.

One of the key components of our design is an efficient modular multiplier, shown in Fig. \ref{ecc_block_diagram}. In order to support arbitrary prime fields, it performs multiplication with interleaved modular reduction \cite{paar_ff_2006}. Three adders are used for this computation, one for addition and two for reduction. The reduction uses conditional subtractions, all performed in the same cycle so that the modular multiplication is constant time and there is no potential timing side-channel. The same circuitry can be re-used for modular addition.

\begin{figure}[!b]
\centering
\includegraphics[width=3.0in]{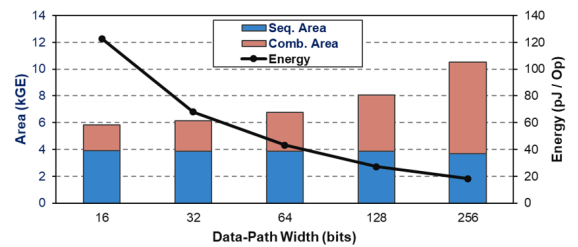}
\caption{Comparison of modular adder architectures, with different data-path widths, in terms of area (sequential and combinational) and energy.}
\label{mod_add_sweep}
\end{figure}

While most ECC designs choose 16-bit or 32-bit data-paths for modular arithmetic, we have used full 256-bit adders for energy-efficiency, with higher bits of the data-path gated when working with smaller primes. Design space exploration was performed for 256-bit modular adders with different data-path sizes, as shown in Fig. \ref{mod_add_sweep}. Clearly, the total area doesn't scale linearly with the data-path width due to the fixed overhead of the 256-bit registers required to store the inputs and output. When scaling up from 16-bit to 256-bit data-path, total area increases by $1.8\times$, while energy per operation decreases by $6.8\times$, primarily due to reduced control circuitry and muxing logic. Table \ref{table:mod_arith_energy} shows the simulated energy consumption of our modular adder and multiplier at 1.2 V and 20 MHz.

Prior work on hardware implementations of ECC re-use the modular multiplier to perform modular inversion using Fermat's theorem: $x^{-1} = x^{p-2}$ mod $p$ \cite{hankerson_ecc_2004}. This method uses repeated modular multiplications (384 on average for 256-bit primes) for exponentiation. Therefore, inversion using Fermat's theorem ($I_{Fermat}$) is slow, but doesn't require any additional logic area. In this design, we make an energy-area trade-off and implement dedicated hardware \cite{banerjee_thesis_2017} to perform modular inversion using the extended Euclidean algorithm ($I_{Euclid}$) \cite{hankerson_ecc_2004}, which involves modular additions, subtractions and bit-shifts. Similar to the multiplier, our inverter also consists of 256-bit adders for energy-efficiency. From Table \ref{table:mod_arith_energy}, energy consumption of the two types of inversions are found to be related to multiplication ($M$) as: $I_{Fermat} \approx 384M$ and $I_{Euclid} \approx 3M$, indicating that $I_{Euclid}$ is $128\times$ more efficient, albeit at the cost of increased logic area.

\begin{table}[!t]
\renewcommand{\arraystretch}{1.2}
\caption{Synthesis results for 256-bit modular arithmetic}
\label{table:mod_arith_energy}
\centering
\begin{tabular}{|l|c|c|}
\hline
\rowcolor{Gray}
\textbf{Operation} & \textbf{Cycles / Op} & \textbf{Energy (nJ / Op)} \\
\hline
\hline
Add / Sub & 1 & 0.02 \\
\hline
Mul & 256 & 4.04 \\
\hline
Inv (Fermat) & 98304 & 1552 \\
\hline
Inv (Euclid) & $\approx$ 720 & 12.9 \\
\hline
\end{tabular}
\end{table}



Having optimized the modular arithmetic implementations, the next step is to select an efficient ECSM algorithm. Traditional window-based ECSM \cite{hankerson_ecc_2004} requires 256 DBL and 64 ADD operations for window-size $w = 4$. Instead, a pre-computation-based comb algorithm \cite{hankerson_ecc_2004, hedabou_zsd_2005} is implemented, which involves 64 DBL and 64 ADD operations, thus reducing ECSM energy by $2.5\times$. A 4 KB cache stores pre-computed comb data for up to six points, including generator points and public keys, which is specifically used to speed up the DTLS handshake, as will be explained in Section \ref{sec:dtls_engine}.

The final optimization step in our design is the appropriate choice of coordinates for elliptic curve points. Resource-constrained ECC implementations \cite{hutter_sardines_2011, ingrid_tinyecc_2013, hutter_tags_2014, hutter_nacl_2015} typically use projective coordinates to avoid modular inversions in the ECSM inner loop, at the cost of extra multiplications and a final expensive Fermat inversion. In projective coordinates, the costs of point opertions are ADD = $8M$ and DBL = $11M$. Since we have an efficient dedicated modular inverter, we use affine coordinates where ADD = $2M + I$ and DBL = $3M + I$. The total ECSM costs of the projective and affine coordinate representations are calculated as $E_{proj} = 64 \times (8M + 11M) + 4M + I_{Fermat} = 1604M$ and $E_{aff} = 64 \times (5M + 2I_{Euclid}) = 704M$. Therefore, the use of affine coordinates saves $\approx 2\times$ in energy by trading off the extra multiplications for cheaper Euclid inversions.

\begin{figure}[!t]
\centering
\includegraphics[width=3.0in]{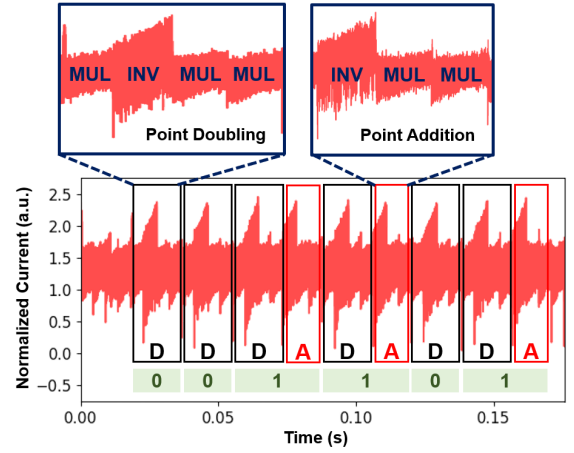}
\caption{Measured power trace demonstrating SPA attack on the simple double-and-add ECSM algorithm implemented in software on the RISC-V processor. The double (D) and add (A) steps are marked, along with their key constituent modular arithmetic operations - multiplication (MUL) and inversion (INV). Also shown are bits of the secret scalar successfully recovered from this trace.}
\label{sw_spa_trace}
\end{figure}

\begin{figure}[!t]
\centering
\includegraphics[width=3.0in]{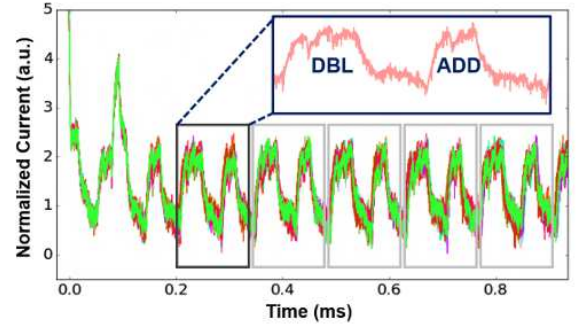}
\caption{Measured power traces of the SPA-secure hardware ECSM, for 10 random scalars, overlaid together for comparison. The sets of point doubling (DBL) and point addition (ADD) operations are shown in boxes, indicating that the double-and-add patterns are constant irrespective of the secret scalar.}
\label{hw_spa_traces}
\end{figure}

Public-key algorithms are prone to side-channel attacks due to their expensive computations and long execution times. One such attack is \textit{simple power analysis} (SPA). Simple double-and-add ECSM algorithms perform conditional point additions in the outer loop depending on whether the corresponding bit in the secret scalar is a 1. Since DBL and ADD involve distinct arithmetic, the power consumption of the chip can leak this information. For reference, we demonstrate an SPA attack on a software implementation of this algorithm, as shown in Fig. \ref{sw_spa_trace}. The slower operations -- multiplication and inversion, can be clearly inferred from a single power trace, and the bits of the secret scalar can be successfully determined. In order to prevent SPA attacks, we use a zero-less signed digit (ZSD) representation of the scalar \cite{hedabou_zsd_2005} in conjunction with the comb technique, which transforms the scalar to have no zero bits, thus avoiding conditional point additions. This also reduces the number of pre-computed comb points per ECSM from 16 to 8. Fig. \ref{hw_spa_traces} shows power traces of our SPA-secure implementation for 10 random scalars overlaid together, where both DBL and ADD are computed at each iteration of the outer loop, irrespective of the bits of the scalar.

The binary scalar $k = (k_{t-1}, k_{t-2}, \cdots, k_1, k_0)_2$ needs to be odd to have a valid ZSD form, that is, the least significant bit $k_0 = 1$ \cite{hedabou_zsd_2005}. To prevent leaking any information about whether $k$ is even or odd, we initially compute $k' = k + 1$ if $k$ is even, and $k' = k + 2$ if $k$ is odd. Then, $Q' = k'P$ is computed, and finally, we obtain $Q = kP$ as $Q' - P$ if $k$ is even, and $Q' - 2P$ if $k$ is odd. We use a compact scalar encoding, which we denote as ZSD$^{*}$, of the ZSD scalar where the 1-bit represents `1' and the 0-bit represents `-1', similar to \cite{joye_encoding_2001}. We prove that this compact form of scalar $k$ can be computed ``on-the-fly'' as ZSD$^{*}(k) = (1, k_{t-1}, \cdots, k_2, k_1)_2$ since the following equation holds: \\
$(1, k_{t-1}, \cdots, k_2, k_1)_2 = $
\[
2^{t-1} + \underbrace{\frac{k-1}{2}}_{\text{+1 bits of } (k_{t-1}, \cdots, k_1)} - \underbrace{(2^{t-1} - 1 - \frac{k-1}{2})}_{\text{-1 bits of } (k_{t-1}, \cdots, k_1)} = k
\]
Therefore, no additional circuitry is required to convert $k$ to the ZSD$^{*}$ form. The SPA countermeasure introduces 5 extra point additions, on average, for 256-bit scalars \cite{hankerson_ecc_2004}, which translates to $\approx 4\%$ energy and performance overhead.

\begin{table*}[!t]
\renewcommand{\arraystretch}{1.2}
\caption{Comparison of our reconfigurable ECC design with state of the art}
\label{table:ecc_comparison}
\centering
\begin{tabular}{|l|c|c|c|c|c|c|}
\hline
\rowcolor{Gray}
\textbf{Design} & \textbf{Tech} & \textbf{Voltage} & \textbf{Logic Area} & \textbf{Supported} & \textbf{Cycles} & \textbf{Energy} \textsuperscript{a} \\
\rowcolor{Gray}
 & \textbf{(nm)} & \textbf{(V)} & \textbf{(kGE)} & \textbf{Curve(s)} & \textbf{/ ECSM} & \textbf{/ ECSM ($\mu$J)} \\
\hline
\hline
Hutter \textit{et al.}, WISTP 2011 \cite{hutter_sardines_2011} \textsuperscript{b} & 350 & 3.3 & 9.5 & NIST P-192 & 753k & 1423.6 \\
\hline
\multirow{2}{*}{Roy \textit{et al.}, ECC 2013 \cite{ingrid_tinyecc_2013} \textsuperscript{b}} & \multirow{2}{*}{32} & \multirow{2}{*}{1.0} & \multirow{2}{*}{26} & SEC P-160 & 250k & 2.25 \\ \cline{5-7}
 & & & & NIST P-192 & 350k & 3.15 \\
\hline
Pessl \textit{et al.}, RFIDSec 2014 \cite{hutter_tags_2014} \textsuperscript{b} & 130 & 1.2 & 8.6 & NIST P-160 & 100k & 4.4\\
\hline
Hutter \textit{et al.}, CHES 2015 \cite{hutter_nacl_2015} \textsuperscript{b} & 130 & 1.2 & 32.6 & Curve25519 & 811k & 56.8 \\
\hline
\multirow{4}{*}{\textbf{This work (Reconfigurable ECC)}} & \multirow{4}{*}{65} & \multirow{4}{*}{0.8} & \multirow{4}{*}{65.5 \textsuperscript{c} (49.1 \textsuperscript{d})} & \multicolumn{3}{c|}{\textbf{All prime curves up to 256 bits}} \\ \cline{5-7}
 & & & & 160-bit & 74k & 2.22 \textsuperscript{e} \\ \cline{5-7}
 & & & & 192-bit & 102k & 3.11 \textsuperscript{e} \\ \cline{5-7}
 & & & & 256-bit & 180k & 6.47 \textsuperscript{e} \\
\hline
\multicolumn{7}{l}{\textsuperscript{a} Base point ECSM energy \textsuperscript{b} Post-synthesis area and power reported in \cite{hutter_sardines_2011, ingrid_tinyecc_2013, hutter_tags_2014, hutter_nacl_2015}} \\
\multicolumn{7}{l}{\textsuperscript{c} Area of final placed-and-routed design \textsuperscript{d} Synthesized area for comparison \textsuperscript{e} Measured energy}
\end{tabular}
\end{table*}


More sophisticated side-channel attacks on ECC exist \cite{ingrid_sca_2010}, which involve statistical metrics, e.g., correlation, and therefore require several power traces for a single scalar. Since the same scalar is never used twice for any of the ECSM computations during the DTLS handshake, these attacks are not particularly relevant to our main application. For other ECC-based protocols, appropriate countermeasures, usually requiring some form of input randomization, can be easily implemented using software.

\begin{figure}[!b]
\centering
\includegraphics[width=3.0in]{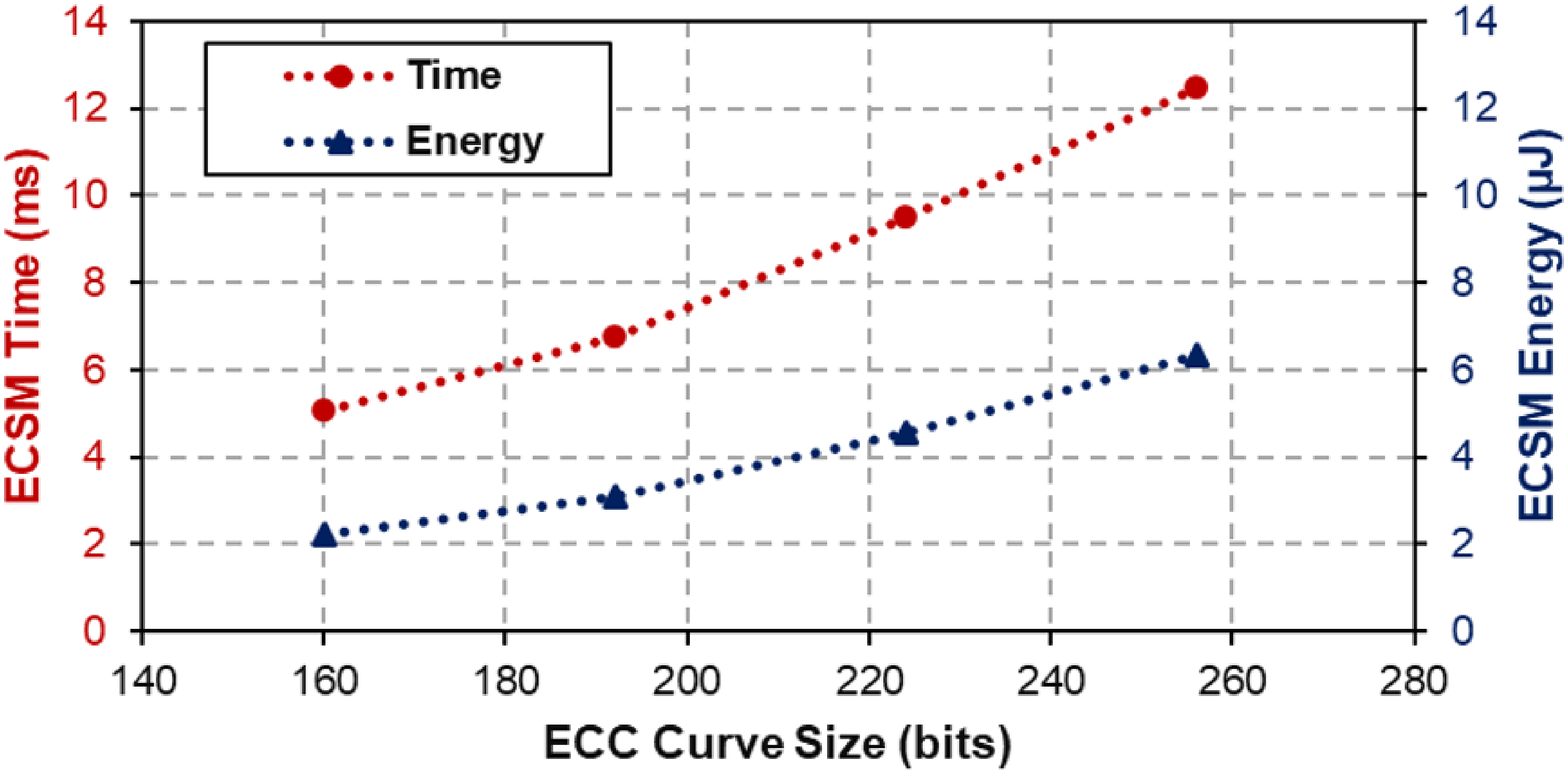}
\caption{Base point ECSM performance over different curve sizes.}
\label{ecsm_perf}
\end{figure}

For a 256-bit short Weierstrass curve, our design takes $\approx 320$k cycles for comb pre-computations and $\approx 180$k cycles for SPA-secure ECSM. Table \ref{table:ecc_perf} compares the measured execution time and energy consumption of our hardware with an equivalent software implementation on the RISC-V at 0.8 V and 16 MHz. As described earlier, our reconfigurable ECC supports all short Weierstrass and Montgomery curves over prime fields up to 256 bits. Fig. \ref{ecsm_perf} shows measured base point ECSM performance over different curve sizes, generated using the NIST curves for 160, 192, 224 and 256 bit primes. ECSM performance and energy scale approximately cubically with size of the prime. Since the prime size is directly related to security, our reconfigurable ECC can be used to scale security and efficiency, depending on the application requirements.

\begin{table}[!b]
\renewcommand{\arraystretch}{1.2}
\caption{Measured 256-bit prime curve ECC performance}
\label{table:ecc_perf}
\centering
\begin{tabular}{|l|c|c|}
\hline
\rowcolor{Gray}
\textbf{Computation} & \textbf{Time (s)} & \textbf{Energy ($\mu$J)} \\
\hline
\hline
Software Comb + ECSM & 8.5 & 4180 \\
\hline
Hardware Comb & 0.02 & 11.1 \\
\hline
Hardware ECSM & 0.012 & 6.47 \\
\hline
\end{tabular}
\end{table}


Our configurable ECC architecture can be easily scaled to larger prime fields (such as 384-bit or 521-bit primes), in order to support security levels higher than 128-bit, by using even wider data-path adders and small changes to the control logic. Table \ref{table:ecc_comparison} compares our design with recent work in ECC hardware. Our design is the most energy-efficient and flexible, but has larger area owing to the dedicated modular inverter (31k GE) and full data-path modular multiplier (11.8k GE). Reconfigurability of our ECC core is also responsible for some of the area overheads, since fixed prime field arithmetic (such as NIST primes) can be implemented with smaller logic with hard-wired parameters.

%% file: body/05_dtls_engine.tex
\section{DTLS Engine}
\label{sec:dtls_engine}

At the core of DTLS is its state machine, which controls all handshaking protocols and related computations. Since the DTLS state machine supports a variety of configurations \cite{rescorla_tls_2018, rescorla_dtls_2018}, software implementations can be error-prone and has lead to attacks in the past \cite{meyer_tls_2013}. To avoid such issues, we enable only a carefully chosen secure subset of all the configurations supported by DTLS. In this work, we have implemented the cipher suite with ECDHE, ECDSA, AES-128-GCM and SHA2-256, requiring mandatory server/client authentication. The \textit{Client Certificate URL} extension is used, that is, client certificates are not transmitted. The \textit{Cached Information} extension is made optional, and the server decides whether to use it during the handshake. Certificate Authority (CA) public keys are cached by both parties, and CA certificates are never exchanged (still maintaining compliance with the TLS specification).

\begin{figure}[!t]
\centering
\includegraphics[width=3.2in]{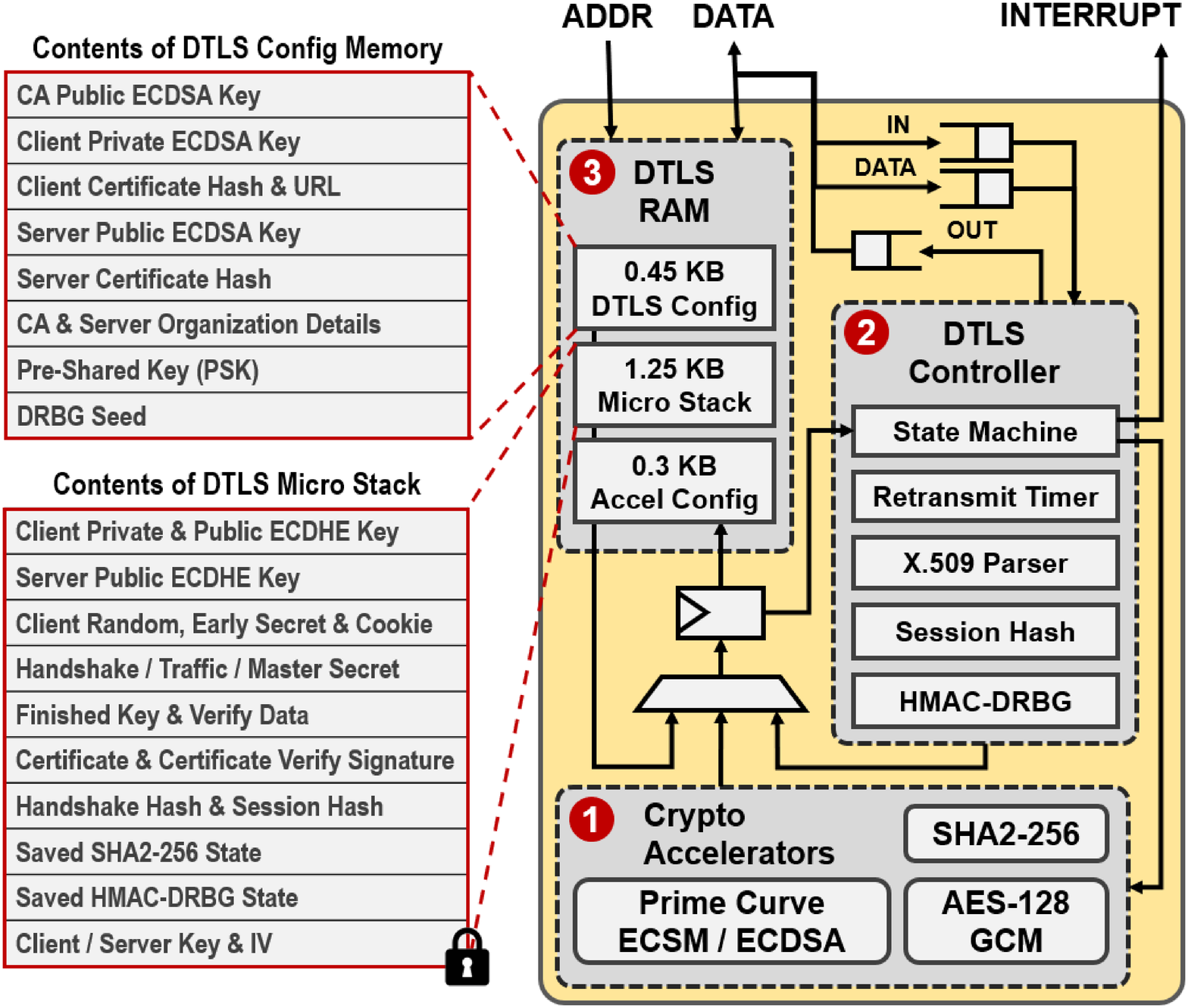}
\caption{Architecture of DTLS engine along with contents of DTLS RAM.}
\label{dtls_core_arch}
\end{figure}

Fig. \ref{dtls_core_arch} shows the architecture of our DTLS engine (DE), with its key components -- (1) energy-efficient cryptographic accelerators, (2) DTLS controller and (3) DTLS RAM. The efficient cryptographic primitives, described in Section \ref{sec:crypto}, not only accelerate DTLS computations but can also be accessed individually through RISC-V software to implement standalone protocols. The DTLS RAM and DTLS controller are discussed in detail in the following subsections.

\subsection{DTLS RAM}

The 2 KB DTLS RAM can be divided into three sections - DTLS micro stack, DTLS Config memory and Accelerator Config memory. The 1.25 KB DTLS micro stack acts as scratch-pad for temporary variables computed during the DTLS handshake, including DRBG states and DTLS session keys. The DTLS stack is not accessible through the memory-mapped interface so that secret session information, including encryption keys, cannot be read by software. The 0.45 KB DTLS Config memory is used to store public keys, secret keys and certificate details, which can be programmed through the RISC-V processor, while the remaining 0.3 KB Accelerator Config memory stores accelerator configuration values for standalone cryptographic operations. Contents of the config memory and the micro stack are detailed in Fig. \ref{dtls_core_arch}.

\subsection{DTLS Controller}

The DTLS controller implements a micro-coded DTLS 1.3 state machine for pseudo-random number generation, key schedule, session transcripts, encrypted packet framing, parsing and validation of X.509 digital certificates and re-transmission timeouts, as shown in Fig. \ref{dtls_core_arch}. Details of some key components of the DTLS controller are discussed next.

\subsubsection{HMAC-DRBG and HKDF}

\begin{figure}[!t]
\centering
\includegraphics[width=3in]{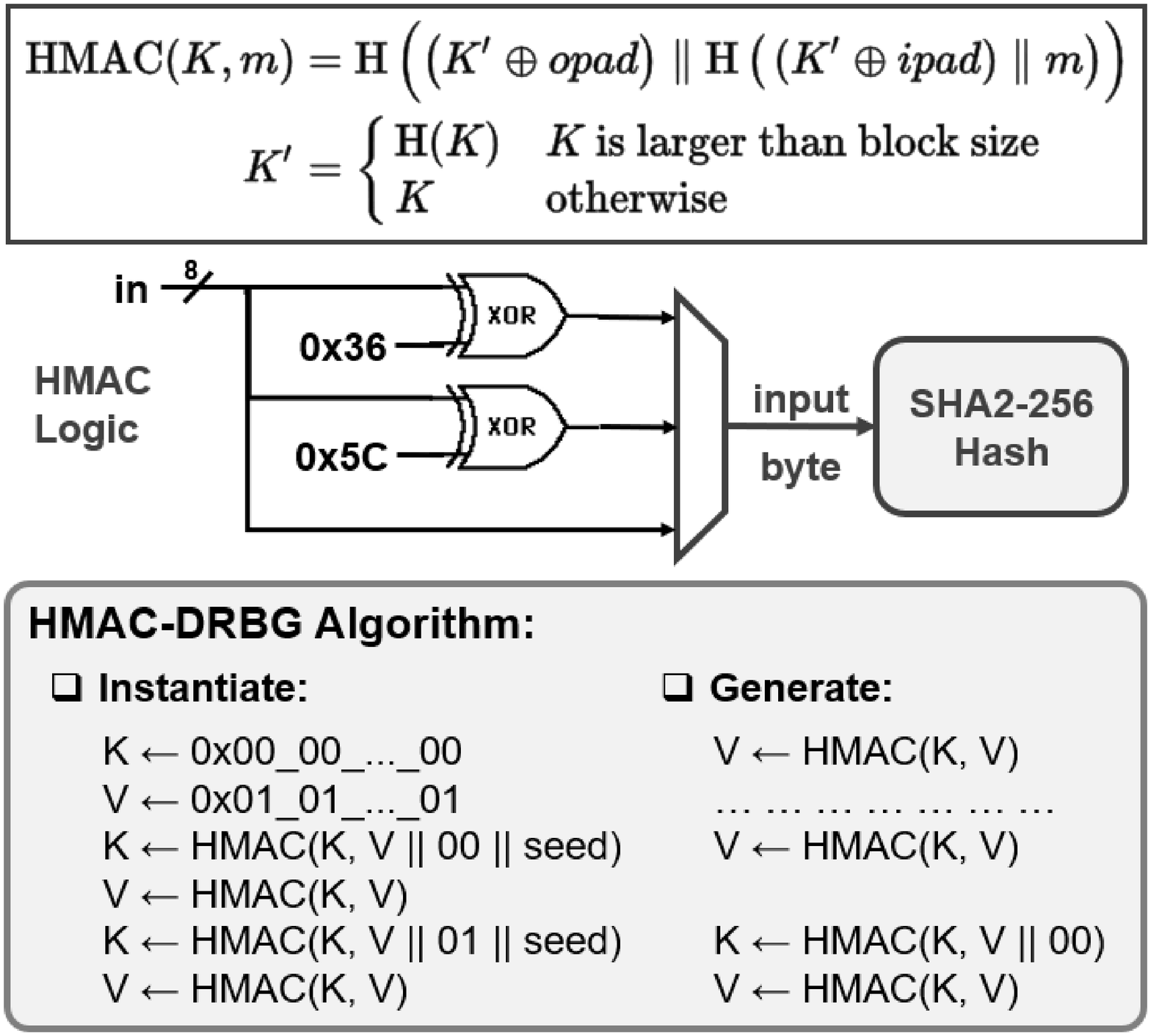}
\caption{HMAC logic along with details of DRBG computations.}
\label{hmac_arch}
\end{figure}

An HMAC-based Deterministic Random Bit Generator (HMAC-DRBG) \cite{nist_drbg_2015} is used to generate cryptographically secure pseudo-random numbers, while an HMAC-based Key Derivation Function (HKDF) \cite{ietf_hkdf_2010} is used to compute DTLS handshake and session keys. Both HMAC-DRBG and HKDF use the SHA2-256 cryptographic accelerator to efficiently compute HMACs (keyed-hash message authentication codes) \cite{ietf_hmac_1997}, as shown in Fig. \ref{hmac_arch}. HMAC uses two passes of the SHA2-256 hash function, with the key XOR-ed with the appropriate \textit{pad} -- repeated bytes valued \texttt{0x36} and \texttt{0x5C} for the inner and outer passes respectively. The HMAC inputs are loaded from the DTLS RAM into a temporary register and bytes are shifted out to the hash module, with or without padding. The micro stack is used to store any intermediate values computed during HMAC.

The HMAC-DRBG algorithm involves two 256-bit values $K$ and $V$, and operates in two phases. The DRBG is first initialized using the input seed material (as obtained from the DTLS Config memory), also known as the \textit{Instantiate} phase. During the \textit{Generate} phase, pseudo-random numbers are generated in $V$, 256 bits at a time, by repeatedly computing HMAC($K$,$V$), followed by an update of $K$ and $V$, as shown in Fig. \ref{hmac_arch}, which are then stored in the micro stack. The DRBG is initialized (seeded) only once at the time of device setup, and it can be used for up to $2^{48}$ invocations of the \textit{Generate} step, with up to $2^{19}$ bits generated in each invocation, as per the NIST DRBG specification \cite{nist_drbg_2015}. Since the DRBG \textit{Generate} function is used 3 times (for generating client random and scalars for ECDHE and ECDSA-Sign) during a DTLS handshake (DRBG not required during application data exchange), it need not be re-seeded for $\approx 9.4 \times 10^{13}$ handshakes, which exceeds the life of the IoT device.

The SHA2-256-based HKDF algorithm also works in two steps -- \textit{Extract} and \textit{Expand}. In the \textit{Extract} phase, a non-secret \textit{salt} and input keying material \textit{IKM} are used to calculate the pseudo-random key \textit{PRK} = HMAC(\textit{salt}, \textit{IKM}). In the \textit{Expand} phase, output keying material is generated in 256-bit blocks $T[k]$ using \textit{PRK} and some application specific \textit{info} as: \\
HKDF-Expand ( \textit{PRK}, \textit{info}, $L$ ) = $T[1]$ $\|$ $T[2]$ $\|$ $\cdots$ $\|$ $T[L/32]$
where $T[k]$ = HMAC ( \textit{PRK}, $T[k-1]$ $\|$ \textit{info} $\|$ $k$ ) for $k > 0$, $T[0]$ = \textit{null}, and $L$ is length of output keying material in bytes.

\begin{figure}[!t]
\centering
\includegraphics[width=3.4in]{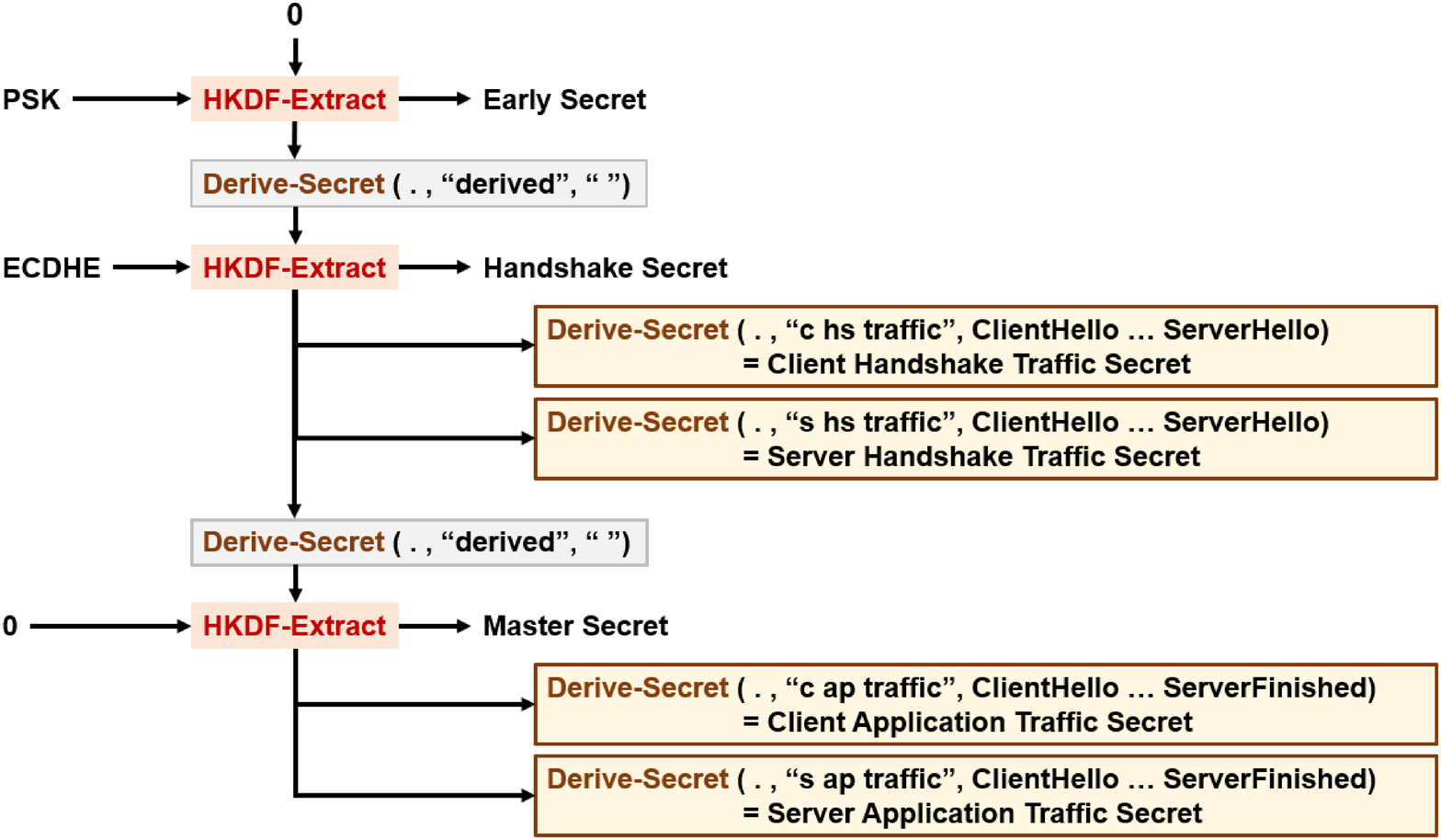}
\caption{TLS 1.3 key schedule \cite{rescorla_tls_2018}.}
\label{key_sched}
\end{figure}

Along with HKDF-Extract, the TLS 1.3 \textit{Key Schedule} \cite{rescorla_tls_2018} uses the following function for key derivation: \\
Derive-Secret ( Secret, Label, Messages ) = \\
HKDF-Expand ( Secret, \texttt{0x0020} $\|$ \texttt{0x746C733133} $\|$ Label $\|$ SHA2-256 ( Messages ), 32 ) \\
The detailed key schedule is shown in Fig. \ref{key_sched}, where PSK refers to the pre-shared key (if PSK is not in use, it is replaced with a string of zero bits) and ECDHE refers to the shared secret computed during the Diffie-Hellman key exchange. The handshake and application traffic secrets are used to generate AES-GCM key and IV pairs (using further invocations of HHKDF-Expand, as specified in \cite{rescorla_tls_2018}) during the handshake and application data phases respectively.

While our prototype chip uses SRAMs and flip-flops as the only on-chip storage elements, a commercial product replicating this design would replace the DTLS config and micro stack with non-volatile memory so that the IoT device can enable power gating while still retaining the configuration values, DRBG state and session keys.

\subsubsection{Session Transcript}

\begin{figure}[!b]
\centering
\includegraphics[width=3.2in]{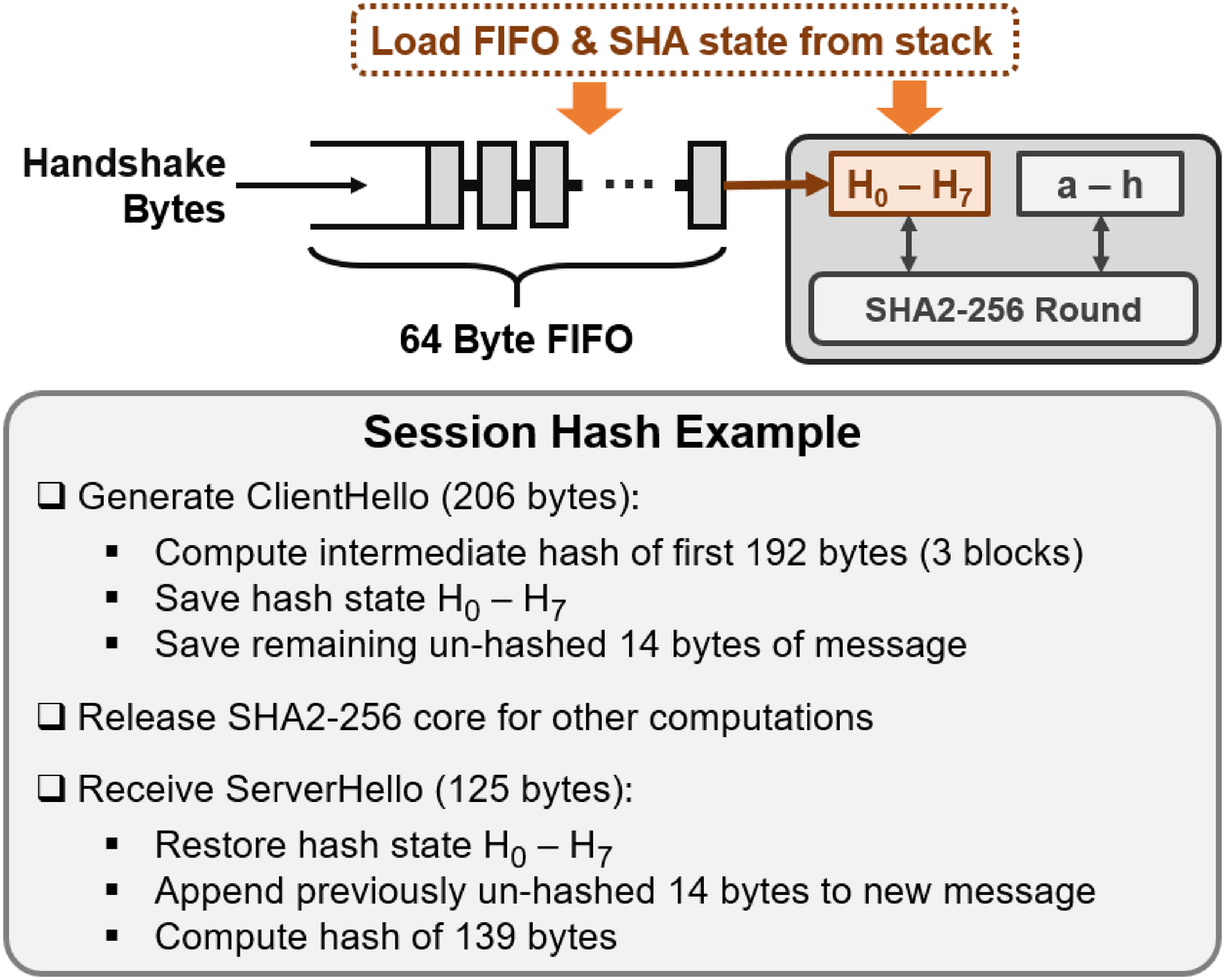}
\caption{Efficient session hash computation for DTLS handshake.}
\label{dtls_session_hash}
\end{figure}

The DTLS handshake involves 6 session hash (transcript) computations, that is, hash of the concatenation of all messages exchanged till that point in the handshake. Software implementations of DTLS typically save all handshake messages, and compute the hash over all of them every time a transcript is required. Handshakes can be as large as 2-3 KB and repeatedly reading them from SRAMs can be very expensive. To eliminate the need to store the entire handshake, we implement a \textit{running hash} by exploiting the property of SHA2-256 that the internal registers $H_0-H_7$ completely define the hash state every time a 512-bit block has been digested, as discussed in Section \ref{sec:crypto}. Fig. \ref{dtls_session_hash} provides an overview of our session hash architecture. Handshake bytes are pushed into a 64-byte FIFO, and a 512-bit block is sent to the SHA2-256 core whenever the FIFO is full. This ensures that session hash computations always digest data in blocks of 64 bytes, except for the last block, and when computing the hash $H(m)$ of an $N$-byte message $m$, the \textit{intermediate hash} of $\lfloor N / 64 \rfloor$ blocks of $m$ is stored in $H_0-H_7$.

In Fig. \ref{dtls_session_hash}, we illustrate this technique using a simple example where the DTLS controller needs to compute SHA2-256 ( \textit{ClientHello} $\|$ \textit{ServerHello} ). After every session hash, the FIFO state (containing any un-hashed bytes) and the registers $H_0-H_7$ are copied to the DTLS stack, so that the SHA2 core can be used for other computations. This is particularly useful for later phases of the handshake which involve hashing large digital certificates. Our proposed approach reduces the total session transcript memory usage from several kilobytes down to only 96 bytes -- 64 bytes for the SHA2 state and up to 32 bytes for the un-hashed portions of the messages.

\subsubsection{ECC Computations in DTLS}

The reconfigurable ECC core is used to perform both ECDH and ECDSA-Sign/Verify computations, where the deterministic ECDSA scheme \cite{pornin_ecdsa_2013} is used to securely generate signatures. The DTLS handshake involves up to 7 ECSM computations, and we have seen in Section \ref{sec:crypto} that ECSM energy can be reduced by $2.5\times$ if pre-computed comb points are available. Our ECC comb point cache supports up to 6 pre-computed base points, which are used to minimize the energy consumption of the ECDHE\_ECDSA handshake. Comb points are computed and stored for the curve generator point $G$, the CA public key $Q_{CA}$ and the server public key $Q_{SRV}$. This one-time pre-computation requires around 33 $\mu$J of energy, which gets amortized over all the subsequent handshakes, but provides up to $2.2\times$ reduction in the energy consumption of each DTLS handshake. The pre-computation for $G$ is essential for both ECDH and ECDSA, while pre-computations for $Q_{CA}$ and $Q_{SRV}$ are used to verify signatures from the CA and the server respectively. The rest of the point cache is used for ECDH and ECDSA with random points without corrupting the stored points required by DTLS.

\subsubsection{DTLS State Machine}

The DTLS state machine is used to generate and process messages at different steps of the handshake as well as exchange encrypted data after the handshake. A 64-bit counter is used to implement the DTLS \textit{Retransmission Timer} \cite{rescorla_dtls_2018} which handles dropped packets. The time-out value can be configured externally, and the state machine re-transmits the previous flight whenever the timer expires. When the DTLS state machine waits for the next flight, all cryptographic accelerators are clock-gated in order to reduce power consumption. Three 256-byte FIFOs are used to fetch input messages (IN FIFO), send output messages (OUT FIFO) and read application data packets (DATA FIFO). The IN FIFO ensures that the DTLS controller starts parsing input messages only when a fully formed packet is available, and sends out complete output messages to the OUT FIFO. For encrypted application data, the state machine also implements the packet optimizations proposed in \cite{banerjee_eedtls_2017}, with the option to enable AES-GCM tag truncation.

%% file: body/06_measurements.tex
\section{Measurement Results}
\label{sec:results}

\begin{figure}[!t]
\centering
\includegraphics[width=3.4in]{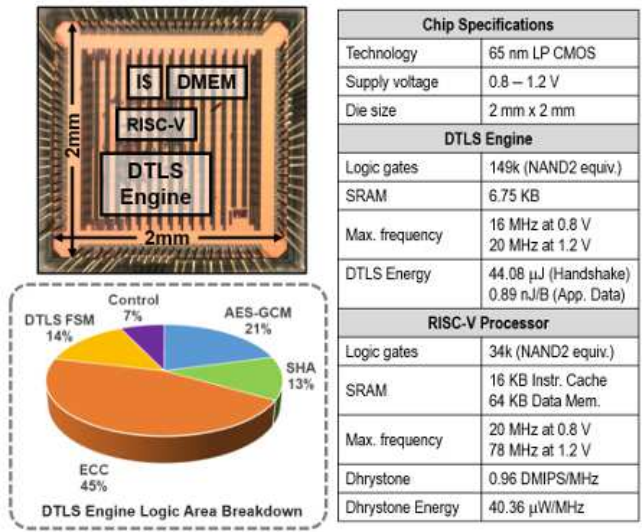}
\caption{Chip micrograph, logic area breakdown of the DTLS engine and summary of chip specifications.}
\label{chip_photo}
\end{figure}

The test chip, shown in Fig. \ref{chip_photo}, was fabricated in a 65 nm CMOS process, with a core size of 1.54 $\times$ 1.54 mm$^2$. The RISC-V processor occupies 0.0489 mm$^2$ (34 kGE) area and interfaces with 16 KB instruction cache and 64 KB data cache. The DTLS engine requires 0.214 mm$^2$ (149k GE) logic area, and uses 6.75 KB of SRAM for the comb point cache, DTLS RAM and packet FIFOs. The chip supports voltage scaling from 1.2 V down to 0.8 V. The RISC-V core achieves a maximum frequency of 78 MHz at 1.2 V and 20 MHz at 0.8 V. The DTLS engine can operate in a frequency range of 16 MHz (at 0.8 V) to 20 MHz (at 1.2 V). All measurements for the RISC-V processor and the DTLS engine are reported at 16 MHz and 0.8 V.

\begin{figure}[!b]
\centering
\includegraphics[width=3.0in]{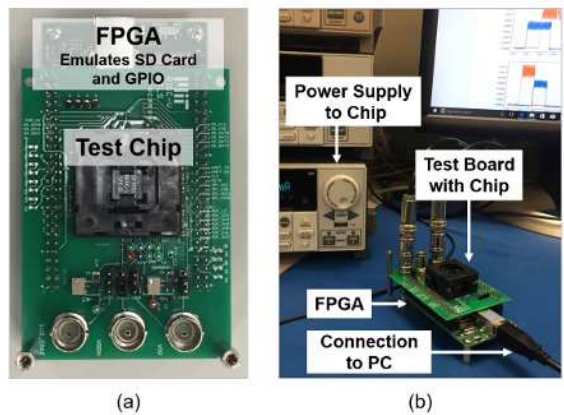}
\caption{(a) Test board with FPGA and (b) power measurement setup.}
\label{test_setup}
\end{figure}

Fig. \ref{test_setup} shows our test board and measurement setup. The test chip is housed in a QFN64 socket soldered to the board, and an Opal Kelly XEM7001 FPGA is used to interface with the chip. A Keithley 2602A source meter is used to supply power to the chip. Both the FPGA and the source meter are controlled from a host computer through USB and GPIB interfaces respectively. While our chip has an SD interface which can communicate with standard SD cards, we use the FPGA to emulate the SD card program memory so that we can eliminate the overhead otherwise imposed by real SD card access times and thus allow fair software benchmarking.

\subsection{Protocol Benchmarks and Energy Measurements}

The DTLS engine supports handshake in two modes:
\begin{itemize}
\item Full -- with verification of server certificate
\item Cached -- with caching of server certificate information to speed up future handshakes
\end{itemize}
The cached mode requires one less ECDSA-Verify operation, thus achieving 36\% reduction in handshake time and energy. Energy consumption of the hardware-accelerated DTLS handshake is 68.94 $\mu$J and 44.08 $\mu$J in the full and cached modes respectively. In the application data phase, the chip consumes 0.89 nJ per byte of data.

\begin{figure}[!t]
\centering
\includegraphics[width=3.0in]{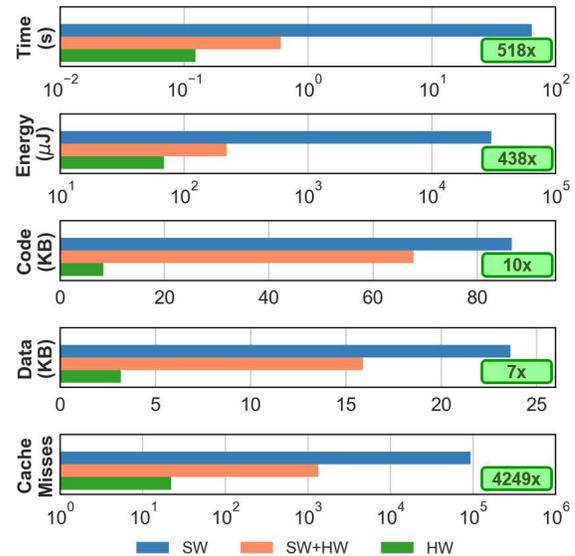}
\caption{Processor resource utilization in three different DTLS handshake implementations -- SW, SW+HW and HW. Improvements of HW versus SW are indicated in the boxes.}
\label{dtls_benchmarks}
\end{figure}

\begin{figure}[!t]
\centering
\includegraphics[width=3.0in]{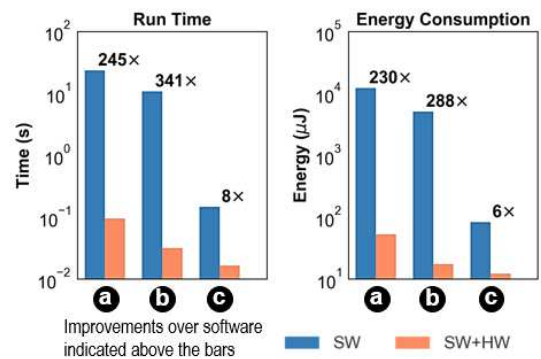}
\caption{Benchmarks for security protocols implemented in SW and SW+HW -- (a) ECMQV, (b) Schnorr Prover and (c) Merkle Hashing. Improvements over software are indicated above the bars.}
\label{protocol_benchmarks}
\end{figure}

In order to analyze the efficiency of our DTLS hardware accelerator, we compared resource utilization in three scenarios: DTLS fully implemented as RISC-V software (SW), the cryptographic kernels accelerated in hardware and only the DTLS controller implemented in software (SW+HW), and DTLS fully implemented in hardware (HW). Test software was implemented using the cryptographic libraries provided by ARM mbedTLS \cite{arm_mbedtls}. Since mbedTLS does not support cached server certificates, all analyses were performed with the DE in non-cached mode. Detailed comparisons are shown in Fig. \ref{dtls_benchmarks}. The use of cryptographic accelerators alone results in over 2 orders of magnitude improvement in run time and energy efficiency (SW vs. SW+HW). The hardware DTLS controller reduces code size by 60 KB, while the DTLS micro stack results in 13 KB reduction in data memory usage (SW+HW vs. HW). When DTLS is accelerated in hardware, code size goes down to only 8 KB, including system functions. We also note that the area occupied by the DTLS state machine and control logic is $5\times$ smaller than the area of SRAM otherwise required to accommodate the DTLS program in software.

Security applications beyond DTLS can also be implemented on the RISC-V, using the cryptographic accelerators in standalone mode. We illustrate this flexibility using three benchmark applications -- (a) ECMQV, an alternative to ECDHE/ECDSA-based authenticated key exchange, (b) Schnorr Prover, an interactive zero-knowledge prover of identity, and (c) Merkle Hashing, used to ensure data integrity in peer-to-peer network protocols. The reduction in resource utilization for all three applications is shown in Fig. \ref{protocol_benchmarks}. The ECC-based applications achieve over $200\times$ increase in energy-efficiency, while Merkle hashing sees $6\times$ energy savings.

Since the cryptographic hardware is active only for a fraction of the total processing time of the IoT node, it is important to analyze the effect of leakage power of the DE on overall energy consumption. Overall leakage power of the chip was measured around 30 $\mu$W at 0.8 V, out of which the DE accounts for $\approx60\%$ (based on layout area). Power consumption of the chip, with only the RISC-V processor active and the DE clock-gated, ranges between 300-650 $\mu$W at 16 MHz and 0.8 V depending on the application software being executed. Therefore, DE leakage power is at most $6\%$ of the total power consumption of the IoT application, thus justifying some of the architectural optimizations described in Section \ref{sec:crypto} which increase leakage power due to larger logic area. The peak current drawn by the chip at 16 MHz and 0.8 V was measured to be around 800-900 $\mu$A, which is well below the maximum current supplied by standard batteries, e.g., 15/30 mA for coin cells.

\subsection{System Demonstration}

\begin{figure}[!t]
\centering
\includegraphics[width=3.2in]{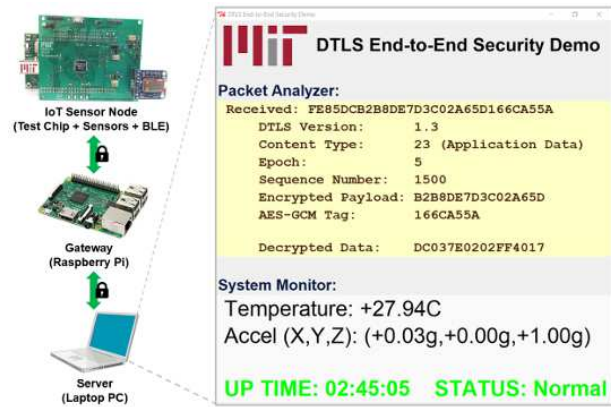}
\caption{System demonstration of a secure IoT node with our test chip collecting data from sensors and transmitting them to a server application over a DTLS-encrypted channel.}
\label{system_demo}
\end{figure}

To demonstrate the functionality of our chip in a complete system, a secure IoT node was designed with the test chip collecting data from a temperature sensor and an accelerometer, encrypting it and then transmitting it through a Bluetooth Low Energy (BLE) transceiver, where all data communications with our test chip are through SPI. A Raspberry Pi module is used as a gateway which forwards these encrypted packets to the application software running on a PC. The system setup is shown in Fig. \ref{system_demo}, along with a screenshot of the server application which displays packet details along with decrypted sensor data.

\subsection{Comparison with Previous Work}

\begin{figure}[!t]
\centering
\includegraphics[width=3.4in]{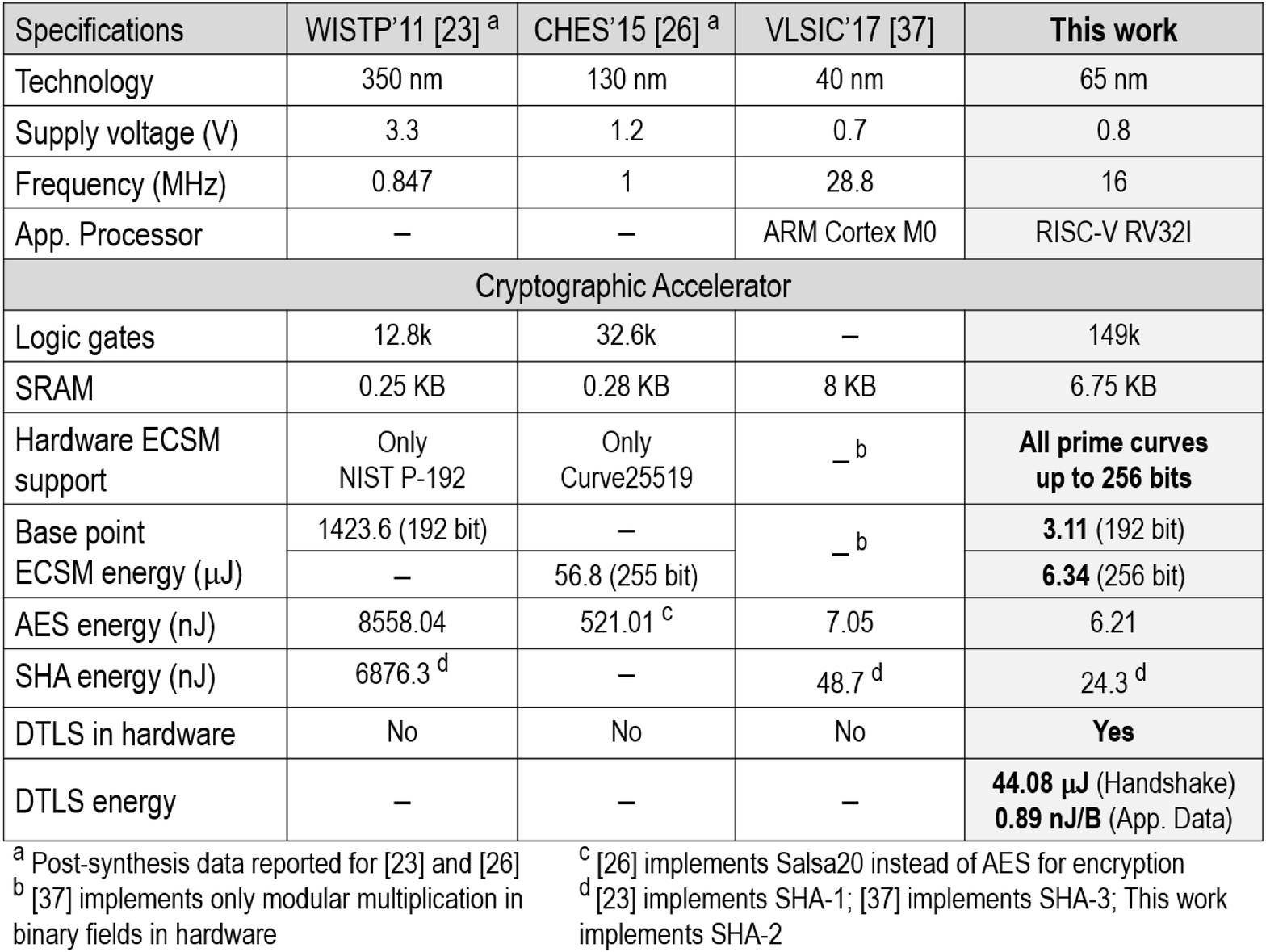}
\caption{Comparison of our design with integrated cryptographic accelerators for embedded systems.}
\label{comparison_table}
\end{figure}

Fig. \ref{comparison_table} compares this work with embedded systems that integrate multiple cryptographic accelerators. This work implements a flexible ECC accelerator which supports arbitrary primes up to 256 bits, in contrast with \cite{hutter_sardines_2011} and \cite{hutter_nacl_2015} which only support fixed 192 and 255-bit curves respectively. \cite{zhang_recryptor_2017} only supports binary field modular arithmetic in hardware. Our ECC accelerator is $458\times$ and $9\times$ more energy-efficient than \cite{hutter_sardines_2011} and \cite{hutter_nacl_2015} respectively at comparable security levels. In addition to the resource savings enabled by the individual cryptographic accelerators, offloading DTLS control flow to the DE realizes a further $3\times$ reduction in energy and $5\times$ reduction in run time compared to a SW+HW implementation.

%% file: body/07_conclusion.tex
\section{Conclusion}
\label{sec:conclusion}

In this work, we have presented an energy-efficient reconfigurable cryptographic engine which makes DTLS a practical solution for implementing end-to-end security on resource-constrained IoT devices. Energy-efficient accelerators for ECC, AES and SHA provide more than two orders of magnitude improvement in performance and energy-efficiency compared to software implementations of DTLS. This allows IoT sensor nodes to re-authenticate more frequently for applications that demand stronger security guarantees. A dedicated DTLS 1.3 protocol controller enables 78 KB and 20 KB reduction in code and memory usage respectively. This allows IoT platforms to implement application programs without having to worry about the overheads otherwise imposed by the security protocol. Protocols beyond DTLS can also be implemented using the RISC-V processor working in conjunction with the cryptographic accelerators, while still getting the benefits of energy-efficiency and performance.

%% file: references.tex


%